\documentclass[preprint,nofootinbib,aps,superscriptaddress,eqsecnum]{revtex4-1}
 \pdfoutput=1
\usepackage{amsmath,amssymb,graphicx,subfigure}

\usepackage{float}
\usepackage{graphicx}
\usepackage{amsmath}
\usepackage{amsfonts}
\usepackage{amssymb}
\usepackage{hyperref}
\usepackage{subfigure}
\usepackage{slashed}
 \usepackage{setspace} 
\usepackage{caption}
\usepackage{color}
\allowdisplaybreaks
\def\bea{\begin{eqnarray}}
\def\eea{\end{eqnarray}}
 \def\be{\begin{equation}}
\def\ee{\end{equation}}
\def\nn{\nonumber}
\def\mpl{M_{\rm Pl}}
\newcommand{\beq}{\begin{equation}}
\newcommand{\eeq}{\end{equation}}

\def\nn{\nonumber}
\def\mpl{M_{\rm Pl}}
\usepackage[utf8]{inputenc}

\DeclareUnicodeCharacter{2212}{-}
\begin{document}

\title{The FIMP-WIMP dark matter in the extended singlet scalar model}

\author{Pritam Das}
\email{prtmdas9@rnd.iitg.ac.in}
\affiliation{ Department of Physics, Tezpur University, Assam-784028, India}
\affiliation{Department of Physics, Indian Institute of Technology Guwahati, Assam 781039, India}
\author{Mrinal Kumar Das}
\email{mkdas@tezu.ernet.in}
\affiliation{ Department of Physics, Tezpur University, Assam-784028, India}
\author{Najimuddin Khan}
\email{najimuddink@ipm.ir}
\affiliation{School of Physical Sciences, Indian Association for the Cultivation of Science
2A $\&$ 2B, Raja S.C. Mullick Road, Kolkata 700032, India}
\affiliation{School of physics, Institute for Research in Fundamental Sciences (IPM), P.O.Box 19395-5531, Tehran, Iran.\vspace{1.80cm}}
\begin{abstract}
We explore the simplest viable dark matter model with a real singlet scalar, vector-like singlet and doublet fermions. The Yukawa couplings associated with the fermion sector play a crucial role in getting the current relic density through Freeze-in and Freeze-out mechanism. We discuss the constraints from the recent muon anomalous magnetic moment experimental data and relic density.
We also perform the collider analysis for the FIMP dark matter in the context of 14 TeV LHC experiments with the MATHUSLA100/200 detector. Our analysis shows that one can get enough events $>3$ for heavy charged fermion track at 14 TeV LHC with an integrated luminosity $\mathcal{L}=10^6 ~{\rm fb^{-1}}$.\end{abstract}

	\maketitle
\section{Introduction}

The standard model (SM) of particle physics is indeed one of the most successful theories of the last century; however, it still has a few unsolved empirical observations. Among those unsolved questions, strong CP problem, neutrino mass and mixing, matter-antimatter asymmetry, nature of dark matter and dark energy have been considered as shortfalls of the SM.  The last few decades have seen an upheaval in astrophysics and cosmology. The Universe is filled with matter and dark matter (DM) and a large amount of dark energy~\cite{Kolb:1990vq}.
From the SM model point of view, there is no sufficient candidate left to propose as a dark matter candidate. Therefore, one must go beyond the SM to explain the current dark matter density.
The recent LHC Higgs signal strength data~\cite{Sirunyan:2017khh, Sirunyan:2018koj} also allows us to include additional fields of the new physics beyond the SM. 

Now one can get the exact relic density of the dark matter using various theories~\cite{Kolb:1990vq, Hall:2009bx}.
Many of the DM genesis theories are based upon the thermal Freeze-out mechanism. In this Freeze-out mechanism, the weakly interacting massive particles (WIMPs) are considered to be in thermal equilibrium in the early universe. 
When the primordial temperature drops below the mass of the DM, $i.e.$, $T<M_{DM}$, it dilutes away until the annihilation and/or co-annihilation to
lighter particles becomes slower than the rate of expansion of the universe; hence, the co-moving DM number density becomes constant.
In the literature, we have now seen that most of the single component WIMP dark matter model are tightly constrained by the recent direct detection limits~\cite{Aprile:2018dbl}. The low mass region may give the relic density in the right ballpark $\Omega h^2=0.1198\pm 0.0026$~\cite{Aghanim:2018eyx}. However, the direct detection limits exclude most of the region~\cite{Burgess:2000yq,Deshpande:1977rw,Khan:2016sxm,Chaudhuri:2015pna}. Hence, multi-component dark matter models are more appealing at the current scenario \cite{Dienes:2011ja}.

Recently the authors of the Ref.~\cite{Hall:2009bx}, have introduced the idea of the Freeze-in mechanism. The dark matter interacts with the other particle feebly, called Feebly Interacting Massive Particles (FIMPs).
Initially, we assume that the dark matter density remains small in the early universe.
The idea is that the dark matter particle(s) get populated through the decay and/or annihilation of the other heavy particles in this model. After a certain temperature, the production of the dark matter through all the processes are ended ($T< M_{Mother~Particles}$), and the co-moving DM number density becomes constant.
It is also confirmed that if the same couplings are involved in both the decay and annihilation scattering processes, then the latter case has negligible contribution to DM relic density over the former one~\cite{Borah:2018gjk,Hall:2009bx,Biswas:2016bfo}.
There are two types of Freeze-in mechanisms that have been discussed in the literature. 
The infra-red (IR) Freeze-in~\cite{Borah:2018gjk,Hall:2009bx,Biswas:2016bfo} and ultra-violet (UV) Freeze-in~\cite{Elahi:2014fsa}. The earlier one based on renormalizable theory and the latter based on theory consists of higher dimension operators in the Lagrangian. The relic density in the UV Freeze-in mechanism depends explicitly on the reheat temperature.
It is also noted that the dark DM density depends on the partial decay width (DM production channels only) of the mother particles. The other decay channel may reduce the mother particles density at $T<M_{DM}$ rapidly. However, at $T>M_{DM}$, the reverse processes can compensate the mother particles density from the other bath particles. But the forward DM production channels from the mother particles  ($p_{mothers} \rightarrow p_{others} \, p_{DM},\, p_{mothers} \rightarrow p_{DM} \, p_{DM}$) are very slow due to the smaller coupling strengths, hence the reverse processes ($ p_{others}\, p_{DM}\rightarrow p_{mothers},\,  p_{DM} \, p_{DM} \rightarrow p_{mothers}$) cannot take place due to the reasons of coupling strengths as well as the initial density of the produce DM particles.
Therefore, the other decay channels of the mother particles will not affect the relic density calculations in the Freeze-in mechanism as the mother particle was thermally equilibrium with the other particles in the early universe~\cite{Hall:2009bx}. Thus, we have to solve only one Boltzmann equation for the evolution of the DM. In this case, one has to consider the sum of all DM production through the decay and annihilation channels of different mother particles.
Suppose the mother particles are not in thermal equilibrium, then we need to solve the evaluation of the mother particles, and at the same time, we have to solve the evaluation for the DM particle~\cite{Biswas:2016bfo}. Before or near the Freeze-in temperature, the mother particles' density should go very small so that the decay and annihilation contributions become zero. Hence, to explain the DM density through the IR Freeze-in mechanism, we need a tiny partial decay of the mother particles, $i.e.$, and the coupling strength should be $\mathcal{O}(10^{-9})$. One may ask the question about the naturalness of the theory, the answer itself lies in the Freeze-in mechanism. By default, we need such a small coupling to explain the DM density. In the UV Freeze-in mechanism, one could achieve these similar coupling strengths by adjusting the heavy cut-off scale in the Lagrangian. However, the dynamic reasons for such tiny coupling strengths are beyond the scope of this work. 

The lepton flavor violating processes ($\mu \rightarrow e \gamma$) along with the muon and electron anomalous magnetic moment are also a striking indication of BSM. The discrepancy between the measured value and the SM predictions is there~\cite{Abi:2021gix, Bennett:2006fi, Parker:2018vye}:
$\delta a_{\mu}=a_{\mu}^\text{exp}-a_{\mu}^\text{SM}=(2.51\pm 0.59)\times 10^{-9}$
and 
$\delta a_{e}=a_{e}^\text{exp}-a_{e}^\text{SM}=-(8.8\pm 3.6)\times 10^{-13}$.
The recent results from Fermilab predicts that the muon anomalous magnetic moment deviates at $4.2\sigma$ from the SM prediction~\cite{Abi:2021gix}. Earlier, it was measured with $3.7\sigma$ deviation from the SM prediction at Brookhaven National Laboratory. The muon magnetic moment is more sensitive hadronic and electroweak contributions, as well as BSM physics due to it's large mass, which influences such larger discrepancies between the $\delta a_{\mu}$. In the meanwhile, the electron anomalous magnetic moment results doesn't allow much deviation\footnote{In this work, we take both the discrepancies as input value to test whether our model can fit those anomalies or not. } \cite{Aebischer:2021uvt}. 
The discrepancy of the muon magnetic moment can be solved by additional Higgs boson\cite{Abe:2017jqo,Chun:2016hzs} and  a light $Z^{\prime}$ gauge boson associated with an extra $U(1)_{L_{\mu}-L_{\tau}}$ symmetry \cite{Baek:2001kca}, or a light hidden photon \cite{Endo:2012hp}, imposing discrete symmetries \cite{Abe:2019bkf}.  Apart from these, various models with collider searches have also been discussed in the literature \cite{Jana:2020joi,Sabatta:2019nfg, vonBuddenbrock:2016rmr,Calibbi:2018rzv,Chen:2021rnl,Yin:2021yqy, Chun:2019sjo}, where both the anomalies and dark matter were discussed nicely.

The addition of the new fields to the SM is widespread in the literature. The lightest and stable particle (at least lifetime of DM should larger than the age of the universe) due to the imposed discrete $Z_n$-charges ($n\ge 2, integer$), behaves as a viable dark matter candidate~\cite{Babu:2009fd}.
Various study on minimal DM models considering scalar and fermion multiplets are available today~\cite{Burgess:2000yq,Deshpande:1977rw,Ma:2008cu,Araki:2011hm,Das:2017ski,Ma:2006km}. In particular, the addition of real singlet scalar, singlet as well as doublet fermion in a minimal model~\cite{Bhattacharya:2017sml,Das:2020hpd}, have rich demand in DM study. The mixing of singlet and doublet fermions reduces the coupling to weak gauge bosons and other fermions, yielding the relic density at the right ballpark with allowed direct detection cross-section~\cite{Cohen:2011ec}. However, most models have discussed the Freeze-out mechanism with WIMP dark matter mass $50$ GeV $- 300$ TeV. The dark matter mass above $300$ TeV violates the perturbativity bound~\cite{Griest:1989wd}, and the low dark matter mass region is either ruled out by the Higgs signal strength data or direct detection search limits.

In the previous work, \cite{Das:2020hpd}, it was found that only the singlet scalar WIMP dark matter model with mass less than $\sim 550$ GeV (except near the Higgs resonance region) are ruled out from recent direct detection limits~\cite{Aprile:2018dbl}. We have also found a considerable increment in the WIMP dark matter parameter spaces in the presence of the new Yukawa couplings. 
One can reduce the Higgs portal coupling $\kappa$ to avoid the direct detection limits.
The additional $t,u$-channels through the new fermions can increase the dark matter annihilation cross-section to produce the exact relic density. 
Regarding the collider search studies, we have found that dilepton$+\slashed{E}_T$ signature can arise from the new fermionic sector and observed at the Large Hadron Collider (LHC).
We have performed the collider analysis in the context of 14 TeV LHC experiments with a future integrated luminosity of 3000 ${\rm fb^{-1}}$ for the final state dilepton$+\slashed{E}_T$ in detail. 
The projected exclusion contour reaches up to $1050-1380~{\rm GeV}$ for 3000 ${\rm fb^{-1}}$ for a light dark matter $\mathcal{O}(10)$ GeV from searches in the $ pp \rightarrow E_1^\pm E_1^\mp, E_1^\pm\rightarrow l^\pm S \rightarrow ll + \slashed{E}_T$  channel. In this study, the SM is extended with a real singlet scalar and singlet and doublet fermion~\cite{Das:2020hpd} is revisited in the context of the low and high dark matter region. 

There have been a rich literature of model studies \cite{Kowalska:2017iqv, Calibbi:2018rzv,Kowalska:2020zve,Kawamura:2020qxo,Baker:2021yli}, where authors have incorporated the analysis of muon anomalous magnetic moments along with neutrino and dark matter mass models. These studies have also tested the validity of their model parameters and dark matter regions in collider searches.
Motivated by these studies, in this minimal setup of our model work, we have identified the new parameter space relevant to low dark matter mass from $10$ keV to $6$ GeV via the Freeze-in mechanism in this present paper. The WIMP dark matter scenario is revisited keeping the collider constraints from the dilepton$+\slashed{E}_T$ searches~\cite{Das:2020hpd}. 
In this minimal model, both low and high DM mass region are studied, including the allowed region from the latest muon anomalous magnetic moment data~\cite{Abi:2021gix} from the Fermilab.
We also carry out the collider study for the FIMP like scenario in the context of 14 TeV LHC experiments with the MATHUSLA100/200 detector. A charged track can be obtained in this model due to the decay of the heavy charged fermion $E_{1,2}^\pm$ into the SM fermions and dark matter. We find that one can get enough events $>3$ at 14 TeV LHC with an integrated luminosity $\mathcal{L}=10^6 ~{\rm fb^{-1}}$.
Soon, if this type of minimal model turns out to be the FIMP and/or WIMP dark matter model realized in Nature, our present (including \cite{Das:2020hpd}) study could estimate a better parameter space.
To the best of our knowledge, a detailed analysis of this model has not been done, which motivating us to study the model. 

The rest of the work is organized as follows. We have given the complete model description in section~\ref{sec2}.
The possible constraints from the lepton flavour violation decay and new results on muon magnetic moment are discussed in section~\ref{sec:lfv}. The dark matter analysis through Freeze-out and Freeze-in mechanism is carried out in detail in section~\ref{dm1}. The possible collider searches from the LHC charged tracker detector is analyzed in section~\ref{sec:collidr}.
Finally, we have conclude our work in section~\ref{conc}. 
\section{Model}\label{sec2}
The model addressed here, contains (i) a real singlet scalar ($S$), (ii) a vector-like charged fermion singlet $E_S^-$ and (iii) a vector-like fermion doublet,
$F_D=(X_1^0~~E_D^-)^T$.
It is to be noted that these additional fermions are vector-like, and hence, they do not introduce any new anomalies ~\cite{Das:2020hpd,pal2014introductory,Pisano:1993es}.
All the newly added particles are considered odd under discrete $Z_2$ symmetry ($\Psi\rightarrow-\Psi$), such that these fields do not mix with the SM fields. Hence, the lightest and neutral particle is stable and can be considered as a viable dark matter candidate. The Lagrangian of the model read as~\cite{Das:2020hpd},
\begin{equation}
	\mathcal{L}=\mathcal{L_{\rm SM}}+\mathcal{L_S}+\mathcal{L_F}+\mathcal{L}_{int},
\end{equation}
where,
\begin{eqnarray}
	\mathcal{L_S}&=&\frac{1}{2}|\partial_{\mu}S|^2-\frac{1}{2}kS^2\phi^2-\frac{1}{4}m_S^2S^2-\frac{\lambda_S}{4!}S^4\label{eq:scalar},\\
	\nn \mathcal{L_F}&=&\overline{F}_D\gamma^{\mu}D_{\mu}F_D+\overline{E}_S\gamma^{\mu}D_{\mu} E_S-M_{ND}\overline{F}_DF_D - M_{NS}\overline{E}_SE_S,\\
	\mathcal{L}_{int}&=&-Y_{N}\overline{F}_D\phi E_S - Y_{fi} \overline{\psi}_{i,L} F_D S  - \, Y_{fi}^{\prime }\, \overline{l}_{i,R} E_S S + h.c.~\label{lint}
\end{eqnarray}
Here, $D_{\mu}$ stands for the corresponding covariant derivative of the doublet and singlet fermions. The left-handed lepton doublet is denoted by ${\psi}_{i,L}=(\nu_i, l_i)_L^T$, whereas $l_{i,R}$ indicate the right-handed singlet charged leptons, with $i=e,\mu,\tau$ three generation of leptons. $L$ and $R$ stand for left and right chirality of the fermions.
The SM Higgs potential is given by,  $V^{SM}(\phi)=-m^2\phi^2+\lambda\phi^4$, with,  
$\phi=( G^+, \frac{H+v+iG}{\sqrt{2}})^T$  is the SM Higgs doublet. $G$'s stand for the Goldstone bosons and $v=246.221$ GeV being the vacuum expectation value of the Higgs $H$ fields. The mass matrix for these charged fermion fields is given by,
\begin{eqnarray}
	\mathcal{M}=\begin{pmatrix}
		M_{ND}&M_X\\M_X^{\dagger}&M_{NS}\\
	\end{pmatrix},
	\label{eq:mass1}
\end{eqnarray}
where, $M_X=\frac{Y_{N}v}{\sqrt{2}}$. The charged component of the fermion doublet ($E_D^\pm$) and the singlet charged fermion ($E_S^\pm$) mix at tree level. 
The mass eigenstates are obtained by diagonalizing the mass matrix with
a rotation of the ($E_D^\pm$  $E_S^\pm$) basis,
\begin{eqnarray}
	\begin{pmatrix}
		E_1^\pm\\E_2^\pm\\
	\end{pmatrix}=\begin{pmatrix}
		\cos\beta&\sin\beta\\-\sin\beta&\cos\beta\\
	\end{pmatrix}\begin{pmatrix}
		E_D^\pm\\E_S^\pm\\
	\end{pmatrix}, {~\rm with~} \tan 2 \beta = \frac{2 M_X}{M_{NS}-M_{ND}}.
\end{eqnarray}
Diagonalization of eqn.~\ref{eq:mass1} gives the following eigenvalues for the charged leptons ($M_{NS}-M_{ND} \gg M_X$) as,
\begin{eqnarray}
	M_{E_1^\pm} = M_{ND} - \frac{2 (M_X)^2}{M_{NS}-M_{ND}},\,
	M_{E_2^\pm} = M_{NS} + \frac{2 (M_X)^2}{M_{NS}-M_{ND}}.\nn
\end{eqnarray}
The masses of the neutral fermion scalar fields can be calculated as,
\begin{eqnarray}
	M_{X_1^0}=M_{ND}, \, M_S^2=\frac{m_S^2+kv^2}{2}~ \ \text{and}~\ M_H^2= 2\lambda v^2. 
	\label{eq:mass}
\end{eqnarray} 
Hence, in this model, neutral fermion can not be the DM candidate as $M_{E_1^\pm}<M_{X_1^0}<M_{E_2^\pm}$. Only the scalar fields $S$ for $M_S < M_{E_1^\pm}$ can behave as a viable DM candidate.
 We will provide a detailed discussion on the new region of the allowed parameter spaces and the effect of the additional $Z_2$-odd fermion in the dark matter section~\ref{dm1}.

The parameter space of this model is constrained by various bounds arising from theoretical considerations like absolute vacuum stability and unitarity of the scattering matrix, observation phenomenons like dark matter relic density. 
The direct search limits at LEP and electroweak precision measurements put severe restrictions on the model.
The recent measurements of the Higgs invisible decay width and signal strength at the LHC put additional constraints.
The dark matter (DM) requirement saturates the DM relic density all alone restricts the allowed parameter space considerably. These constraints are already discussed in our previous paper~\cite{Das:2020hpd}. We discuss the lepton flavour violation ($\mu\rightarrow e\gamma$), incorporating the new anomalous magnetic moment result in the next section.

\section{Lepton flavor violation ($\mu\rightarrow e\gamma$) and anomalous magnetic moment}\label{sec:lfv}
In this model, the LFV couplings (in our model, $Y_{fi}$ and $Y_{fi}^{\prime}$ with $i=e,\mu,\tau$) and model parameters got severely constraint by the bounds from LFV processes. Hence, it is evident that LFV bounds can mimic other observable like DM parameter spaces (one can go through the WIMP dark matter analysis section, carried out in this work).
The observed dark matter abundance is typically obtained for $\kappa=Y_{fi}^{all}=\mathcal{O}(0-1)$ through $s$-channel,  $t$-channel annihilation and the combination of these two processes (co-annihilation, $i.e.$, mass differences can also play a crucial role). The lepton flavour observable are expected to give stringent constraints on the parameter spaces.

\begin{figure}[h]
	\centering
	\includegraphics[scale=.65]{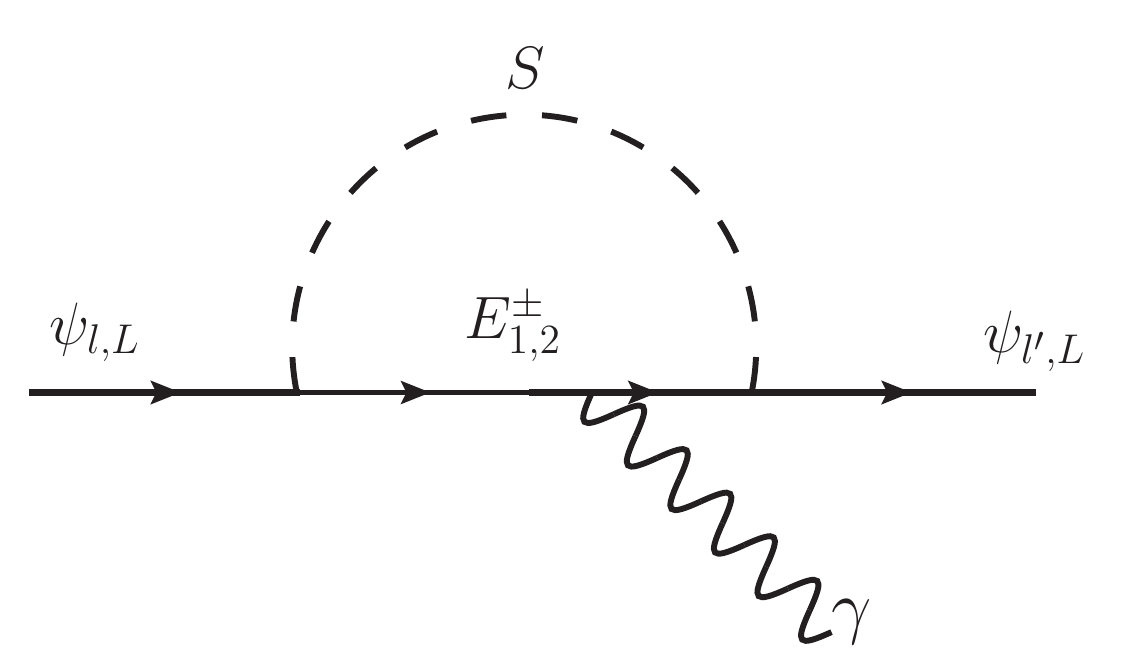}
	\includegraphics[scale=.65]{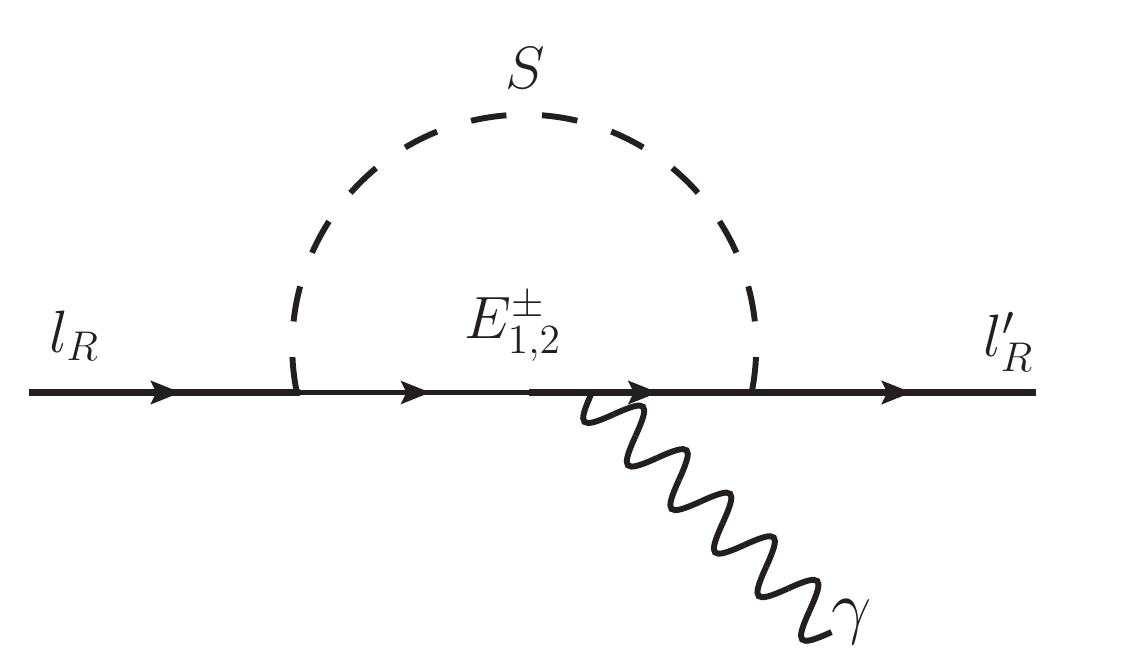}
	\caption{\it  Muon and electron anomalous magnetic moment and LFV process $\mu\rightarrow e\gamma$ decay diagrams mediated by charged particles $E_1^{\pm}$ and $E_2^{\pm}$. }\label{lfv}
\end{figure}

Among the various LFV processes, the radiative muon decay
$\Gamma(\mu\rightarrow e \gamma)$ is one of the popular and restrictive one, which in the present model is mediated by charged particles $E_1^\pm, E_2^\pm$ present in the internal lines of the
one-loop diagram~\ref{lfv}. 
The corresponding expression for the branching ratio is given by,
\begin{eqnarray}
	{\rm BR} (\mu\rightarrow e \gamma) = \frac{3 \alpha_{em}}{64 \pi G_F^2} \Big | A^\dagger_{f1} A_{f2} \frac{ F(M_{E_{1}^\pm}^2/M_S^2) }{ M_S^2}+ B^{\dagger}_{f1} B_{f2}\frac{ F(M_{E_{2}^\pm}^2/M_S^2) }{ M_S^2}\Big |^2 ,
	\label{eq:fl}
\end{eqnarray}
where, $
F(x)=\frac{x^3-6 x^2 +3 x+ 2 + + x ln(x)}{6 \, (x-1)^4}
$.
The coupling strengths of the corresponding vertices are $A_{f1}= \cos\beta Y_{f1} + \sin\beta Y_{f1}^{\prime} $ $ A_{f2}= \cos\beta Y_{f2}  + \sin\beta Y_{f2}^{\prime}$, $B_{f1}= \sin\beta Y_{f1}  + \cos\beta Y_{f1}^{\prime}$ and $ B_{f2}=\sin\beta Y_{f2}  + \cos\beta Y_{f2}^{\prime}$ respectively.
With the most recent experimental bounds for LFV~\cite{Baldini:2018nnn},
one can get allowed parameters from the lepton flavor violating decay constraints~\cite{Baldini:2018nnn} $ {\rm BR} (\mu\rightarrow e \gamma) < 4.2 \times 10^{-13}$ at $90\%$ CL., for $Y_{fj}$ and $Y_{fj}^{\prime}<\mathcal{O}(10^{-3})$ with $j=e$ or $\mu$.

Along with the LFV process, the muon anomalous magnetic moment provides new sensitivity for new physics contribution due to the higher level of precision in both theoretical and experimental measurements. The recent measurement of the muon anomalous magnetic moment, $a_{\mu}=(g−2)_{\mu}/2$, at Fermilab shows a discrepancy $w.r.t.$ the Standard Model (SM) prediction \cite{Abi:2021gix}, 
\begin{eqnarray}
	a_{\mu}^{FNAL}=116592040(54)\times10^{-11},\\
	a_{\mu}^{SM}=116591810(43)\times10^{-11},
\end{eqnarray}
and, when it was combined with the previous Brookhaven result \cite{Bennett:2006fi},
$
	a_{\mu}^{BNL}= 116592089(63)\times10^{-11}$
 leads to an observed excess of $\delta a_{\mu}=251(59)\times10^{-11}$ at $4.2\sigma$ C.L. 
In the near future experiments at Fermilab, these discrepancies are expected to be clarified with more precise data.
The electron anomalous magnetic moment sensitivity is~\cite{Abi:2021gix, Bennett:2006fi, Parker:2018vye}: 
$\delta a_{e}=a_{e}^\text{exp}-a_{e}^\text{SM}=-(8.8\pm 3.6)\times 10^{-13}$.
It is negative and one need a very sensitive parameters to solve the electron as well as muon anomalous magnetic moment simultaneously.
In our model, with the vector-like fermions in hand, the new contribution to anomalous magnetic moment can be written as~\cite{Chao:2008iw,Chen:2019nud},
\begin{eqnarray}
	\delta a_i &=& \frac{ Re[A_{fi}^2] m_l^2 }{8 \pi^2}  \int_0^1 \, \frac{ x (1-x)^2}{(x-x^2) m_l^2  + (x-1) M_{E_1^\pm}^2 - x M_{S}^2 }dx \nn\\
	& & + \frac{ Re[B_{fi}^2] m_l^2 }{8 \pi^2}  \int_0^1 \, \frac{ x (1-x)^2}{(x-x^2) m_l^2  + (x-1) M_{E_2^\pm}^2 - x M_{S}^2 }dx,
	\label{g2}
\end{eqnarray}
where $i=e,\mu,\tau$. As we can see directly from equation \eqref{g2}, the contribution to $\delta a_i$ depends on the masses ($m_{E_{1,2}^\pm}, M_S$), Yukawa couplings ($Y_{fi}$ and $Y_{fi}^{\prime}$) and the mixing angle ($\beta$). 
Depending on the new Yukawa couplings $Y_{fi}$ and $Y_{fi}^{\prime}$, the loop diagram simultaneously contributes to the electron as well as muon anomalous magnetic moment.  For simplicity, let us assume all the $Y_{fi}^{\prime}$ couplings are negligibly small.
We find that, a very large Yukawa couplings ($Y_{fi} > 4\pi$) are required to explain the present lepton anomalous magnetic moment data. For example, one can get $\delta a_\mu=2.51\times10^{-11}$ for $Y_{f2} =29.32$, $M_S=1000$ GeV, $M_{E_1^\pm}=1500$ GeV and  $M_{E_2^\pm}=3000$ GeV. This choice of couplings and masses violate the LFV as well as perturbative limits. 
If we assume non-zero $Y_{f2}=Y_{f2}^{\prime}$, then we also need $Y_{f2}\sim \mathcal{O}(20)$ to get $\delta a_\mu=2.51\times10^{-11}$. Similarly, we also need a very large $Y_{f1}=Y_{f1}^{\prime}$ to get electron anomalous magnetic moment.
On the contrary, these large couplings are not healthy for this model as it will render a large negative potential (become unbounded from below) towards the singlet as well as the Higgs scalar fields due to radiative corrections~\footnote{The detailed analysis with radiative corrections is beyond scope of this work}.  
To work out this model smoothly, we always kept $Y_{f2} $ and $Y_{f2}^{\prime}<\mathcal{O}(10^{-3})$.
It is to be noted, the regions allow by LFV data are also allowed by these anomalous magnetic moment data.
In the next section, we discuss the dark matter analysis and bound coming from the relic density.
\section{Dark matter}\label{dm1}
As pointed out in the previous section, the viable DM candidate in this model is the lightest $Z_2$-odd singlet scalar $S$. Here, the dark matter can give exact relic density through the Freeze-out mechanism and/or Freeze-in mechanism, depending on the choice of parameter spaces. Suppose the dark matter is in thermal equilibrium in the early universe; in this case, the dark matter can annihilate to the  SM particles when $T>M_{DM}$, where $T$ is the temperature of the universe. It Freezes out for $T<M_{DM}$, and depending on the parameter spaces it could give the exact relic density. However, if it is not in thermal equilibrium in the early Universe, it could produce from some mother particles and give correct relic density through the Freeze-in mechanism.
Dark matter resulting from a decay or annihilation of various mother  particles, is in thermal equilibrium at early universe. This condition is given by, 
\beq
{\Gamma \over H(T) } \geq 1,
\eeq
where, $\Gamma$ is the relevant decay width and  $H(T)$ is the Hubble parameter given by~\cite{Plehn:2017fdg,Hall:2009bx,Biswas:2016bfo}
\beq
H(T) = \left( g^* \, \frac{\pi^2}{90} \, \frac{T^4}{\mpl^2} \right)^{1/2},
\label{eq:Hub}
\eeq
where, $\mpl=2.4\times 10^{18}$ GeV is the reduced Planck mass and $T$ is the temperature ($1~ \text{GeV}=1.16 \times 10^{13}~ \text{Kelvin}$).
If the production of mother particles occur mainly from the annihilation of other particles in the
thermal bath, $\Gamma$ will be replaced by~\cite{Plehn:2017fdg,Hall:2009bx,Biswas:2016bfo}
\beq
\Gamma = n_{eq} <\sigma v>,
\eeq

where, $n_{eq}$ is their equilibrium number density and is given by~\cite{Plehn:2017fdg}
\beq
\begin{aligned}
	&n_{eq} &=
	&\left\{ \begin{array}{l} \vspace{0.3cm} g^* \left( \frac{m T}{2 \pi}\right)^{3/2} \, e^{-m/T},  ~~~~~~~{\rm for ~non\text{-}relativistic ~states}~~T<<M\\
		
		\frac{\zeta_3}{\pi^2} g^* T^3,  ~~~~~~~~~~~~~~~~~~~~{\rm for~relativistic~boson ~states}~~T>>M\\
		
		\vspace{0.5cm}
		\frac{ 3}{4}\, \frac{\zeta_3}{\pi^2} g^* T^3 ,  ~~~~~~~~~~~~~~~~~~{\rm for~relativistic~fermion ~states}~~T>>M
		\vspace{-0.2cm}\end{array}
	\right.
	\label{eq:n}
\end{aligned}
\eeq
where, the Riemann zeta function has the value $\zeta_3=1.2$ and $g^*$ is the effective degrees of freedom in this framework. Here,
$<\sigma v>$ is the thermally averaged annihilation cross-section of the particles in the thermal bath and can be expressed as~\cite{Gondolo:1990dk,Plehn:2017fdg} 
\beq
<\sigma_{xx} v> = \frac{  2 \pi^2 T \, \int_{4 m^2}^\infty ds \sqrt{s} \, (s-4 m^2)  \, K_1(\frac{\sqrt{s}}{T})  \sigma_{xx}   }{   \left( 4 \pi m^2 T K_2(\frac{m}{T})  \right)^2        },
\eeq

where, $K_{1,2}$ is themodified Bessel function of functions of order 1 and 2 respectively. The DM production through annihilation depends upon the Higgs portal couplings $\kappa$ through $s$- and cross-channels (one can reverse the Figs.~\ref{fig:DarkAn}-(a), ~\ref{fig:DarkAn}-(b) and ~\ref{fig:DarkAn}-(c)) and the new Yukawa coupling $Y_{fi}$ through $t$-channel ( Figs.~\ref{fig:DarkAn}-(d)). Earlier we have checked \cite{Das:2020hpd} that for $\kappa, Y_{fi} \sim \mathcal{O}(0.001)$ with DM mass $\sim \mathcal{O}$(GeV), the dark matter is in thermal equilibrium at early Universe, $i.e.$, $\frac{ n_{eq}<\sigma_{xx} v>}{H(T)} >> 1$. These dark matter mass region give the exact relic density via the Freeze-out mechanism.

In this work, we find that the non-thermally produced singlet scalar can also serve as a viable dark matter candidate at $\mathcal{O}(1)$ GeV (depends on the parameters $\kappa$, $Y_{fi}$ and $\beta$) satisfying the dark matter relic density in the right ballpark. Those light dark matter can interact with other particles very feebly.  
For such very weakly interacting particles, called feebly interacting massive particles or FIMPs, one can invoke the non-thermal dynamics, so-called the Freeze-in mechanism. This mechanism needs weak interactions, which could be one reason to have a tiny fine-tuned coupling in this model. In this model, we find that the shallow dark matter mass region $\mathcal{O}(1)$ keV-MeV could also give the exact relic density in the right ballpark.

We will discuss both the dark matter regions coming from the Freeze-out and Freeze-in mechanism in the following two subsections keeping eye on all the constraints. We want to refer to our previous paper~\cite{Das:2020hpd} for the detailed constraints arising from various theoretical and experimental point of view.

\subsection{WIMP dark matter}
\begin{figure}[h!]
	\begin{center}
		\subfigure[]{
			\includegraphics[scale=0.6]{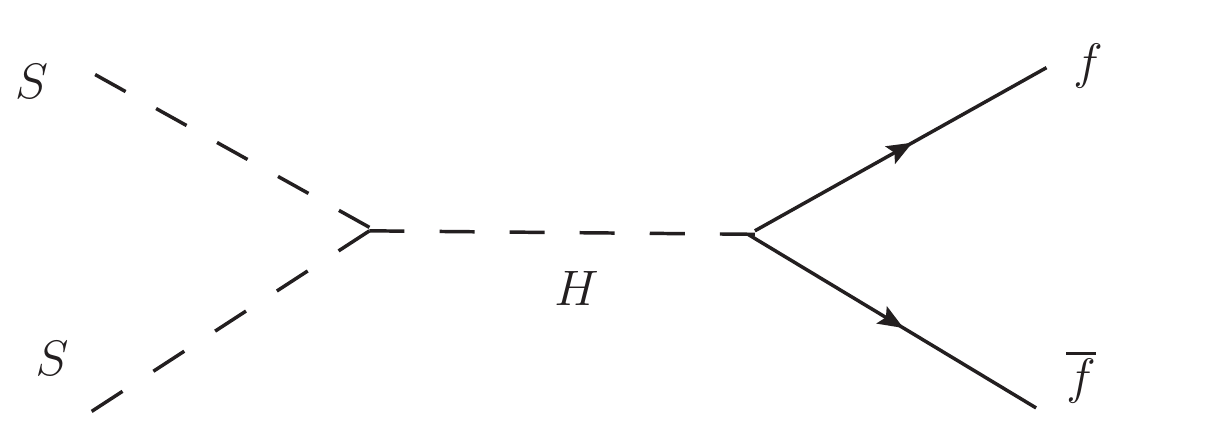}}
		\hskip 1pt
		\subfigure[]{
			\includegraphics[scale=0.6]{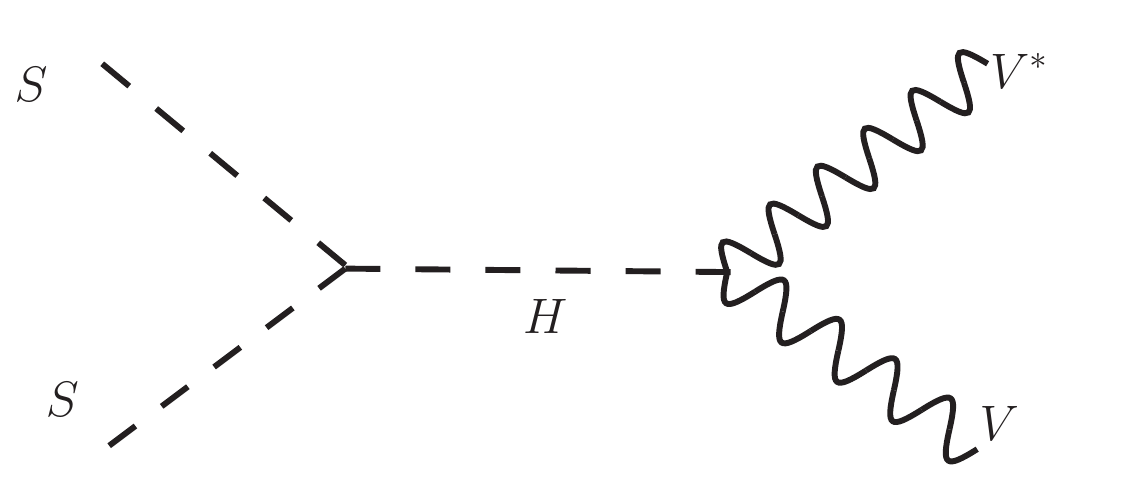}}
		\subfigure[]{
			\includegraphics[scale=0.6]{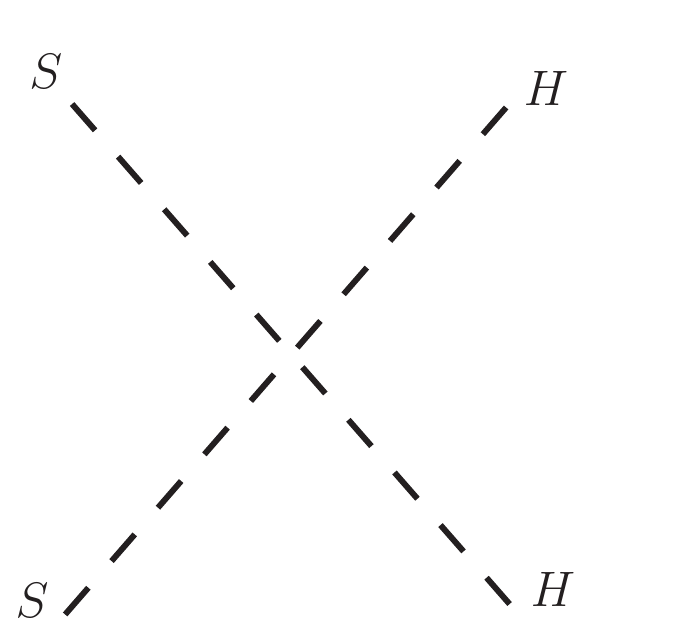}}
		\hskip 1pt
		\subfigure[]{
			\includegraphics[scale=0.6]{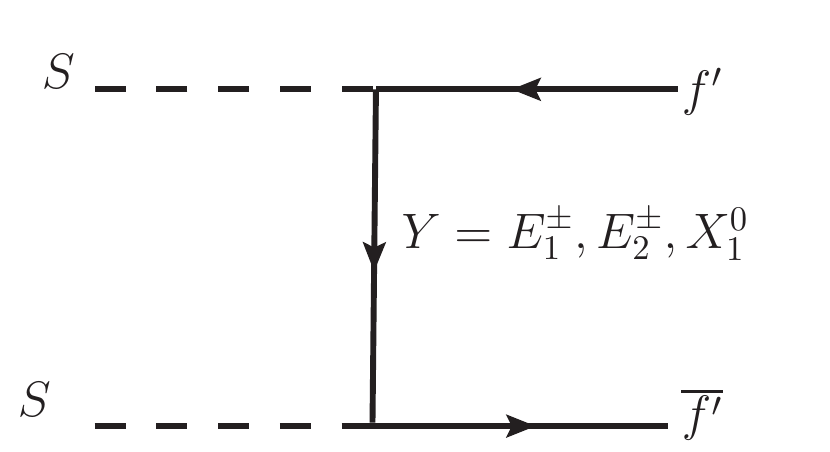}}
		\caption{ \it The DM annihilation diagrams give the relic density. $V$ stands for gauge bosons $W,Z$; $f'$ represents the SM leptons and $f$ are SM leptons and quarks.}
		\label{fig:DarkAn}
	\end{center}
\end{figure}

The lightest $Z_2$-odd singlet scalar $S$ plays the role of viable WIMP dark matter candidate for the choice of region of the parameter spaces with mass $\mathcal{O}(100)$ GeV and annihilation coupling strength $\mathcal{O}(0.1)$~\cite{Das:2020hpd}. 
In our WIMP dark matter study, we use {\tt FeynRules}~\cite{Alloul:2013bka} to get the input codes for {\tt micrOMEGAs}~\cite{Belanger:2018mqt} and compute the relic density of the scalar DM. We also verified the results using {\tt SARAH-4.14.3}~\cite{Staub:2015kfa} including {\tt SPheno-4.0.3}~\cite{Porod:2011nf} mass spectrum in {\tt micrOMEGAs} and get the same relic density of the scalar DM.
The Higgs portal couplings $\kappa$ controls the DM production and/or annihilation through $s$- and cross-channels (see figs.~\ref{fig:DarkAn}-(a), ~\ref{fig:DarkAn}-(b) and ~\ref{fig:DarkAn}-(c)).
The Yukawa couplings associate with the singlet scalar ($Y_{fi}$ and $Y_{fi}^{\prime}$) and the charged fermions ($Y_N$) also have significant influences~\cite{Das:2020hpd} in the WIMP DM parameter space. Depending on the Yukawa couplings ($Y_{fi}$ and $Y_{fi}^{\prime}$), DM can annihilate via $t$- and $u$-channels (see fig.~\ref{fig:DarkAn}-(d)) in our model.
The relevant vertices for the dark matter annihilation and/or co-annihilation are given by:
\begin{eqnarray}
g_{HSS}&=& | \kappa v|,  ~g_{HHSS}= |\kappa|,\nn\\
g_{SiE_1^\pm}&=&A_{fi}=  |(\cos\beta Y_{fi} + \sin\beta Y_{fi}^{\prime})| ,\nn\\
g_{SiE_2^\pm}&=&B_{fi}= |(\sin\beta Y_{fi} + \cos\beta Y_{fi}^{\prime})|,\\
g_{S\nu_i X_1^0}&=&C_{fi}= | Y_{fi} , ~{\rm with}~i=e,\mu,\tau|. \nn
\label{eq:annivertex}
\end{eqnarray}

\begin{figure}[h!]
	\begin{center}
		\subfigure[]{
			\includegraphics[scale=0.6]{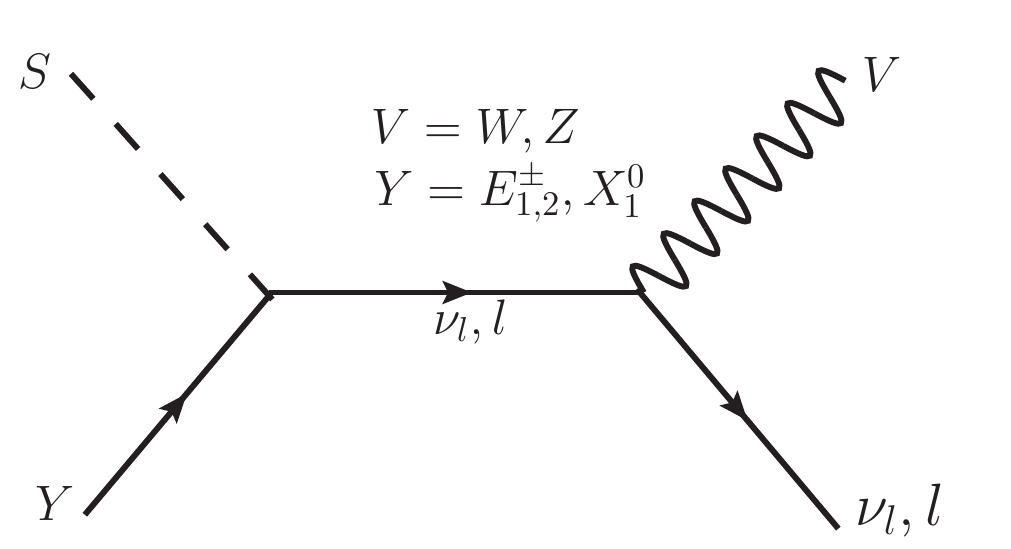}}
		\hskip 1pt
		\subfigure[]{
			\includegraphics[scale=0.6]{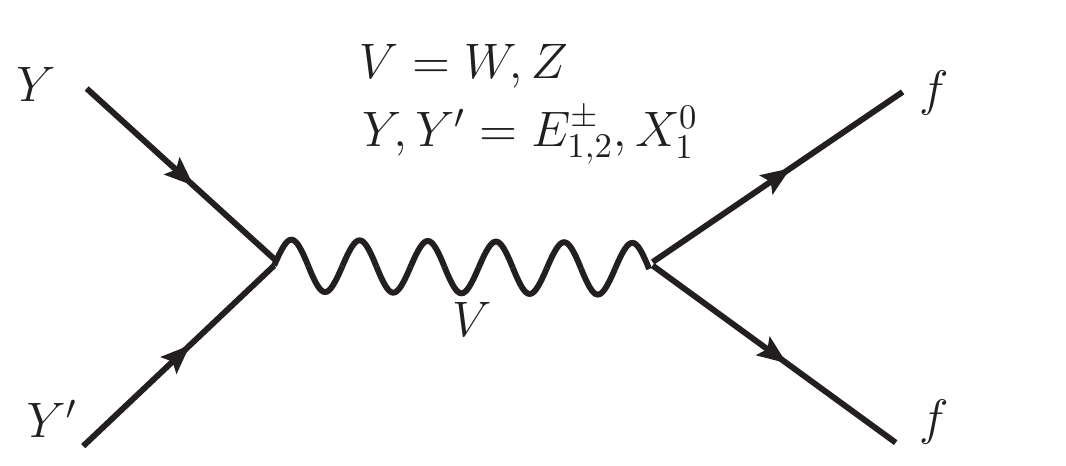}}
		\subfigure[]{
			\includegraphics[scale=0.6]{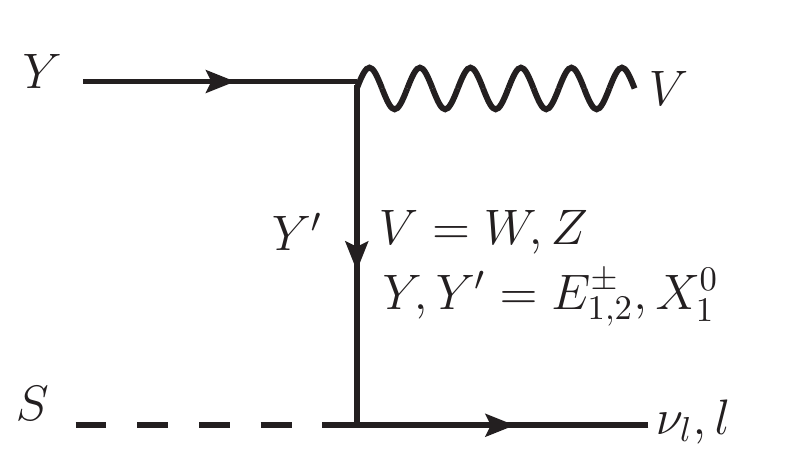}}
		\caption{ \it The co-annihilation and annihilation diagrams of the DM and the other $Z_2$-odd fermion fields. $f$ are SM leptons and quarks.
		}
		\label{fig:DarkCoan}
	\end{center}
\end{figure}

The interference between the $s$-channel, cross-channel and $t,u$-channels played a crucial role to achieve the correct DM density. 
The co-annihilation channels (e.g., see Fig.~\ref{fig:DarkCoan}) also have an essential role in getting a viable region of allowed dark matter parameter space.
The contributions from the $S$ mediate co-annihilation $t,u$-channel diagrams is negligibly small. If the mass difference between the DM and other $Z_2$-odd
particles are within $2\%-10\%$, it is expected that co-annihilation will dominate over DM self-annihilation~\cite{Griest:1990kh}.
We find a huge suppression to the relic density due to large co-annihilation for $M_{E_{1,2}^\pm, X_1^0}-M_S\approx 50$ GeV. Hence most of the higher mass region are ruled out due to the under abundance of a single component dark matter.
It is to be noted that the Sommerfeld enhancement~\cite{ArkaniHamed:2008qn} do not play any role to enhance the current dark matter phenomenology due to $M_{E_{1,2}^{\pm}}>M_{DM}$.

We already have seen that dilepton$+\slashed{E}_T$ signature can arise from the new fermionic sector through $ pp \rightarrow E_1^\pm E_1^\mp, E_1^\pm\rightarrow l^\pm S \rightarrow ll + \slashed{E}_T$  channel.
The charged fermion may observe at the Large Hadron Collider (LHC) with high luminosity.
We have already done such analysis in our previous work~\cite{Das:2020hpd} in the context of 14 TeV LHC experiments with a future integrated luminosity of 3000 ${\rm fb^{-1}}$.
By performing a detailed cut based collider analysis, we have seen that a large region of the parameter spaces can be probed/excluded by the LHC experiments.
The projected exclusion contour have reached up to $1050-1380~{\rm GeV}$ for 3000 ${\rm fb^{-1}}$ for a light dark matter $\mathcal{O}(10)$ GeV from searches in the $ pp \rightarrow E_1^\pm E_1^\mp, E_1^\pm\rightarrow l^\pm S \rightarrow ll + \slashed{E}_T$  channel.
Hence, in this analysis, we keep fixed the mass parameters $M_{E_1^{\pm}}=1500$ GeV and $M_{E_{2}^{\pm}}=3000$ GeV. 
For simplicity, we consider the effect from the interaction term $- \, Y_{fi}^{\prime }\, \overline{l}_{i,R} E_S S$  (see eqn.~\ref{lint}) to be zero, with $Y_{fi}^{\prime}=0$. We will discuss the effect at the end of the WIMP dark matter section.
We also assume $Y_{f1}=Y_{f3}=Y_f$ with $Y_{f2}=0.001$~\footnote{ In general, large Yukawa coupling are also allowed from the relic density. One can also get similar results for the choice of $Y_{f2}=Y_{f3}=Y_f$ and $Y_{f1}=0.001$ or $Y_{f2}=Y_f$ with $Y_{f1}=Y_{f3}=0.001$. For different choice of couplings,  Fig. 4 (left) remains almost identical; however, the right one gets modified due to having only one dominant contribution from the diagram with the second Yukawa coupling.} to avoid flavor violating decays and fixed mixing angle as $\cos\beta=0.995$. This small mixing angle ($\beta$) diluted the contribution from the second charged fermion $E_2^\pm$.
We now scan a prominent region for the DM mass in this work, Higgs portal coupling $\kappa$ and new Yukawa coupling $Y_f$.
The dark matter mass $M_{DM}$ have been scanned from $\sim 5$ GeV to 1500 GeV with a step size of $5$ GeV while Higgs portal coupling changes from $-0.6$ to $0.6$ with step size $0.001$ and new Yukawa coupling from $-0.5$ to $0.5$ with step  $0.001$. 

\begin{figure}[h!]
	\centering
	\includegraphics[scale=.34]{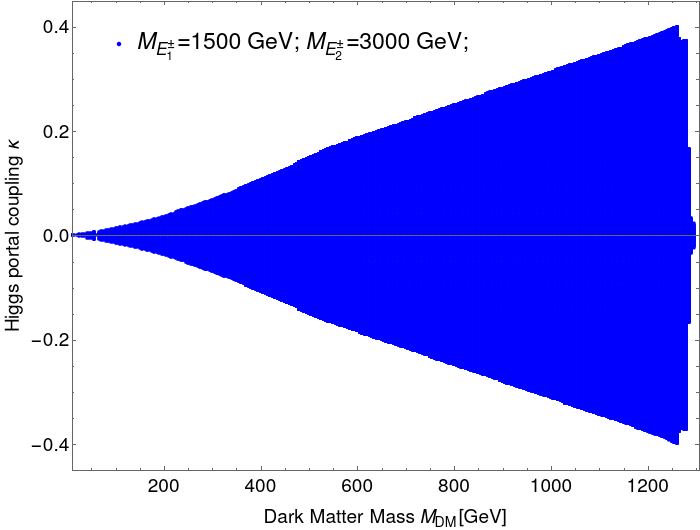}
	\includegraphics[scale=.34]{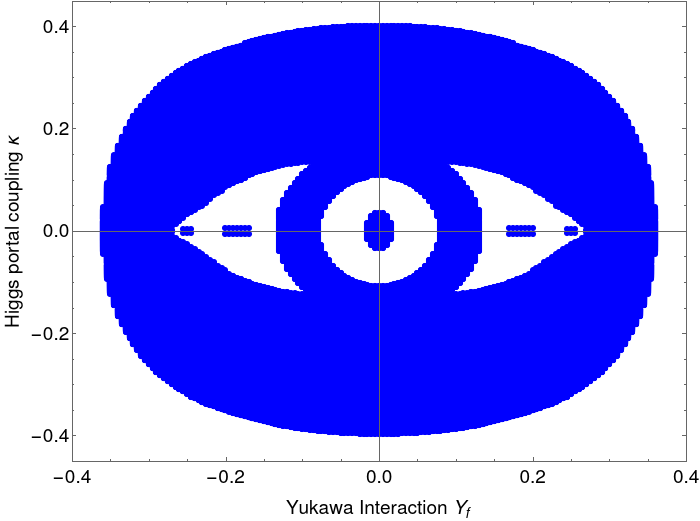}
	\caption{\it  The blue points indicate the relic density within the $3\sigma$ range. These plot are generated by varying the dark matter mass, Higgs portal coupling $\kappa$ and the new Yukawa coupling $Y_f$. The empty region is in the second plot are mostly disfavoured by the relic density and direct detection constrainsts.  The parameters allowed by relic density are also allowed by the recent LFV ~\cite{Baldini:2018nnn} BR$(\mu \rightarrow e\gamma) $ $< 4.2 \times 10^{−13}$ at $90\%$ CL, electron as well as muon anomalous magnetic moment $g-2$ data at Fermilab~\cite{Abi:2021gix}.}	\label{fig:relicall}
\end{figure}

For $\Delta M^{\pm,0}<0.1 M_{DM}$~\cite{Griest:1990kh} ($\Delta M^\pm= M_{E_1^\pm}-M_{DM}$ and $\Delta M^0= M_N-M_{DM}$), the co-annihilation channels (Fig.~\ref{fig:DarkCoan}\footnote{ The DM does not participate in diagram 3(b),  but this diagram plays an essential role in relic density calculation through co-annihilation processes when $M_{Y}-M_S\approx (2-10)\%$ of $M_S$.}) play an important role in the dark matter density calculation\footnote{ There is a missing diagram corresponding to $S f_{V}\rightarrow f_{SM} H$, mediated by the t-channel exchange of a $f _V$, however,  due to propagator-vertex suppression, the annihilation channel have very tiny contribution, therefore we have not included that diagram}. It is noted that the co-annihilation effects are completely absent here as $\Delta M^{\pm,0}> 0.1 M_{DM}$. We display the allowed parameters $\kappa-M_{DM}$ plane in Fig.~\ref{fig:relicall}(left). In the shallow mass region (below $M_H/2$), $t,u$-channel annihilation processes play a key role in giving rise to correct relic bounds. Those regions are also allowed by Higgs decay width and direct detection bounds. $SS\rightarrow \nu\nu$ are the main dominant channels in this region. In the low DM mass region, annihilation via $s$-channel, cross-channel and other co-annihilation processes are negligibly small. The Higgs portal coupling is kept very small ($\kappa\sim 0$) to avoid the Higgs signal strength constraints. On the other hand, all the processes, depending on the parameters in the high mass region, are important to provide the exact relic density. The above and below region correspond to Higgs portal coupling $|\kappa| \gtrsim \frac{M_{DM}}{3324.92~{\rm GeV}}$ of \ref{fig:relicall}(left) are strictly ruled out from the direct detection data.

In the absence of new fermion interaction ($Y_f=0$), DM-mass region in between 70 GeV to 450 GeV is ruled out by the present direct detection cross-section~\cite{Aprile:2018dbl}. However, we do not have any DM signature in those experiments at the current date, and by adjusting Higgs portal coupling $\kappa$ and the new Yukawa coupling $Y_f$, one can work out this forbidden region. 
As we increase the viable allowed DM mass, the Higgs portal coupling $\kappa$ keeps increasing if we neglect the Yukawa coupling $Y_f$.
Similarly, if we choose a small Higgs portal coupling $\kappa$, one has to increase the Yukawa coupling $Y_f$ with DM mass to get the relic density at the right ballpark. Variation of these couplings can exceedingly increase dark matter parameter spaces through the combination of  $s$-,$t$- and $u$- annihilation and co-annihilation channels.

\begin{table*}[h!]
	\centering
	\begin{tabular}{|p{1.6cm}|p{1.2cm}|p{1.1cm}|p{1.2cm}|p{1.2cm}|c|p{4.7cm}|}
		\hline
		\hline
		Channel & $M_{DM}$ (GeV) & ~~$\kappa~~~$& $M_{E_1^\pm}$ (GeV) &~~$Y_{f}$&$\Omega_{DM}h^2$&~~~~~~~~~~Percentage \\
		\hline

		&&&&&&$\sigma(S S\rightarrow \nu \nu)\quad~98\%$\\
			~~BP-1&10&0.00&1500~~&0.29&0.115& $\sigma(SS \rightarrow  ll)\quad 2 \%$\\
		\hline
		&&&&&&$\sigma(S S\rightarrow \nu \nu)\quad~98\%$\\
		~~BP-2&80&-0.01&1500~~&0.28&0.1142& $\sigma(SS \rightarrow  ll)\quad 2 \%$\\
		\hline				&&&&&&$\sigma(S S\rightarrow \nu\nu)\quad~46\%$\\
		&&&&&&$\sigma(S S\rightarrow W^{\pm}W^\mp)\quad~23\%$\\
		~~BP-3&270&0.058&1500~~&0.24&0.113& $\sigma(SS \rightarrow  HH)\quad 13 \%$\\
		&&&&&&$\sigma(S S\rightarrow ZZ)\quad~11\%$\\
		&&&&&&$\sigma(S S\rightarrow ll)\quad~7\%$\\
		\hline
				&&&&&&$\sigma(SS\rightarrow W^\pm W^\mp)\quad 48\%$ \\
		~~BP-4&760&0.224&1500& 0.09 &0.1243& $\sigma(SS\rightarrow HH) \quad24\%$\\
				&&&&&&$\sigma(S S\rightarrow ZZ)\quad~24\%$\\
						&&&&&&$\sigma(S S\rightarrow ll)\quad~3\%$\\
		\hline
		&&&&&&$\sigma(SS\rightarrow W^{\pm} W^{\mp})\quad 49\%$ \\
		~~BP-5&1000&0.31&1500& 0.05 &0.114&$\sigma(SS\rightarrow ZZ) \quad24\%$\\
		&&&&&& $\sigma(SS\rightarrow HH)\quad24\%$\\
		&&&&&& $\sigma(SS\rightarrow ll)\quad2\%$\\
		\hline
		&&&&&&$\sigma(E_1^\pm S\rightarrow W^{\pm} \nu)\quad 45\%$  \\
		~~BP-6&1285&0.372&1500& 0.035 &0.112& $\sigma(X_1S\rightarrow W^\pm l) \quad 32\%$ \\
		&&&&&& $\sigma(SS\rightarrow \nu \nu)\quad9\%$\\
		&&&&&& $\sigma(SS\rightarrow W^\pm W^\mp)\quad2\%$\\
		\hline
	\end{tabular}
	\caption{\it The benchmark points allowed by all the theoretical and experimental constraints. $\sigma(SS\rightarrow \nu \nu)$ is mainly dominated by the $t+u$-channel annihilation processes whereas $\sigma(SS\rightarrow YY),\,Y=W,Z,H,t$ dominated by the  $s+cross$-channel annihilation processes. We consider $Y_{f1}=Y_{f3}=Y_f$ to avoid flavor violating decay processes.}
	\label{tabDM:3}
\end{table*}

The $SS\rightarrow VV, HH$ (with $V=W^\pm, Z$) channels keep on dominating in the $mid-high$ (500-1000 GeV) DM mass regions. Various allowed benchmark points are shown in Table \ref{tabDM:3}. 
One can see, in the presence of DM annihilation via $s$-channel and $t,u$-channels, as most of the regions are giving the correct DM density, which are also allowed by other experimental constraints. For $\kappa\sim 0$, the $s$-channel annihilation still dominates near Higgs resonance region $\sim \frac{M_H}{2}$. The other mass regions give an overabundance for the same $\kappa$ and $Y_f\sim 0$. For a small $\kappa \sim 0$, the $t+u$-channels contributes to get the correct relic density at $3\sigma$ C.L., whereas $s$-channel and $cross$-channel processes dominate for large $\kappa$ and small $Y_f$.
We have shown the viable region of parameters in the $\kappa-Y_f$ plane in Fig.~\ref{fig:relicall}(right).  A blue circular eye-like pattern is obtained here. The same data points are also shown in the $\kappa-M_{DM}$ plane of Fig.~\ref{fig:relicall}(left). The blue regions are the allowed data points passes by all experimental and theoretical bounds~\cite{Das:2020hpd} such as  stability, perturbativity, unitarity, LHC diphoton signal strength, electroweak precision experiments, lepton flavour violation and lepton anomolus magnetic moments\footnote{ We are getting $\delta\alpha_i<10^{-15}$ (with $i=e,\mu$) hence, lepton anomalous moment bounds are satisfied here. On the contrary, to get exact value of $\delta\alpha_i$, we need very large Yukawa coupling, which is restricted by the perturbativity bounds.}. The empty region violates one of the constraints, such as the relic density of the dark matter, direct detection, and Higgs decay width for the DM mass $<\frac{M_H}{2}$. There is large excluded region for $\kappa,Y_{f}\sim0$ compared to the presence of both $\kappa$ and $Y_f$.  
For example, the points $Y_{f}\approx 0, \kappa\approx 0$ give exact relic density for the dark matters masses near Higgs resonance, i.e., $55<M_{DM}<64$ GeV. We get under abundance around $M_{DM} \approx M_h/2\pm 1$. This can be seen in the left panel of the Fig. 4.
The central region of the eye is also dominated mainly by the co-annihilation channels where $\Delta M^{\pm,0}<0.1 M_{DM}$ for high dark matter masses. The following circular empty region is produced under abundance due to a large co-annihilation cross-section for the high dark matter mass. The blue iris-eye region is now allowed for the other masses, where the co-annihilation effects are negligible. Both the side of this iris-eye region is also produced under-abundance due to the effect of additional dark matter annihilation through the $s,t$ and $u$ channel processes for different dark matter masses. The small blue region (both the left and right side) gives the exact relic density for the low dark matter masses $M_{DM}<100$ GeV through the $t$ and $u$ channel processes. The bigger blue region has all possible effects on the relic density for various dark matter masses. The outer side of the blue region is mostly ruled out from the direct detection data.
The low DM mass region was satisfying the current relic density within 3$\sigma$ C.L., obtained via the $t,u$-channel processes. 
The $s$-channel annihilation processes in the low mass region were considerably absent; hence, very few allowed points for $\kappa\sim0$ were observed on the right side of Fig. \ref{fig:relicall}. In those scenarios mainly $E_1^\pm E_1^{\pm}\rightarrow YY$~(with $Y=W^\pm,Z,H)$ via $t,u$-channels and other annihilations and/or co-annihilations dominate and the relic density gets under-abundant for large $Y_f$ and become overabundant for $Y_f=0$. As we move towards the higher mass regions, both the annihilations and co-annihilations processes start dominating. Please note that, the co-annihilations processes fully dominate in the region where we have $\Delta M^{\pm,0}<< 0.1 M_{DM}$. Hence, large points satisfying the relic density within 3$\sigma$ C.L. are observed for a wider variety of $\kappa$, $Y_f$ and $M_{DM}$.

\begin{figure}[h!]
	\centering
	\includegraphics[scale=.3]{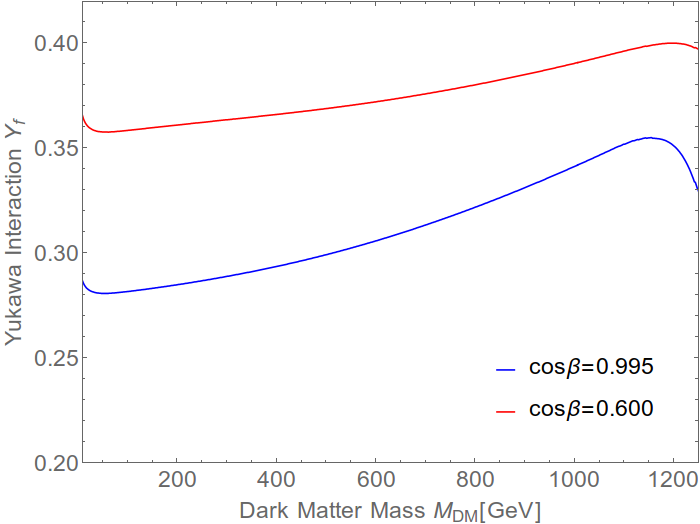}
	\includegraphics[scale=.3]{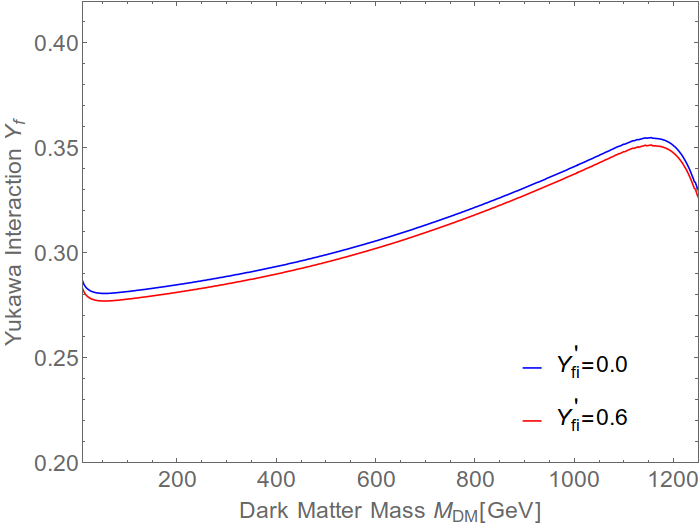}
	\caption{ The blue and red lines indicate the relic density $\Omega_{DM}h^2=0.1198$. These plots are generated by varying the dark matter mass and the new Yukawa coupling $Y_f$ for different $\cos\beta$ and $Y_{fi}^\prime$.
 The values $Y_{fi}^\prime=0$ (left panel) and $\cos\beta=0.995$ (right panel) remain fixed. 
The masses $M_{E_1^\pm}=1500$ GeV and $M_{E_2^\pm}=3000$ GeV and  $\kappa=0$  are same for both cases. These parameters allowed by relic density are also allowed by the recent LFV, electron as well as muon anomalous magnetic moment $g-2$ data. }\label{fig:yfpeffect}
\end{figure}

It is important to highlight that the small mixing angle ($\beta$) and large mass $M_{E_2^\pm}=3000$ GeV diluted the contribution from the new second charged fermion $E_2^\pm$. This eigenstate is mostly composed of the $SU(2)$ singlet charged fermion $E_S^\pm$ due to this choice of small mixing angle.
	One can increase the percentage of $E_D^\pm$ component in the same eigenstate $E_2^\pm$
	by decreasing the value of this mixing angle. At the same time the coupling strengths of the $t,u$-channels vertices $g_{E_{1}^\pm l^\mp S}=\cos\beta Y_{fi}$ and $g_{E_{2}^\pm l^\mp S}=\sin\beta Y_{fi}$ get modified. It influences the annihilation of the dark matter through $t,u$-channels; hence the relic density will change. For example, one can find the exact relic density allowed by all the other constraints for the dark matter mass $M_{DM}=270$ GeV (see BP-3 of table~\ref{tabDM:3}) with Higgs portal coupling $\kappa=0.058$ and $Y_{f}=0.25$ with fixed $\cos\beta=0.995$ and $M_{E_2^\pm}=3000$ GeV. In this case, if we change $\cos\beta=0.6$, then we get an overabundance of the dark matter density $\Omega_{DM} h^2>0.15$. To achieve the exact relic density, we need to decrease the value of the mass of the second eigenstate and/or increase the new Yukawa coupling $Y_f$.
	For example, by lowering the mass eigenstate of $E_2^{\pm}$ to  $M_{E_2^\pm}=1510$ GeV and keeping the same $Y_f=0.24$ or keeping $M_{E_2^\pm}=3000$ GeV with $Y_f=0.301$, one can get  $\Omega_{DM} h^2=0.113$. It is important to note that the left plot of the Fig.~\ref{fig:relicall} will remain almost same for the choice of different mixing angle $\cos\beta$ and fixed $M_{E_2^\pm}=3000$ GeV as the Higgs portal coupling $|\kappa|$ is constrained from the dark matter direct detection. However, the right plot of Fig.~\ref{fig:relicall} will be modified. One can have larger allowed region in the $Y_f-\kappa$ plane from the relic density and other constraints. The allowed values of new Yukawa coupling can reach up to $Y_f\leq 0.391$ for $\kappa=0$ and $\cos\beta=0.6$ in the $Y_f-\kappa$ plane. We show the effect of the mixing angle in Fig.~\ref{fig:yfpeffect} (left) of the $t,u$-channels contribution in the relic density.  

One can also consider non-zero $Y_{fi}^{\prime }$ to get enhancement (for the same sign of $Y_{fi}$) in the dark matter annihilation as well as co-annihilation into the SM particles, and for $\cos\beta=0.995$ this enhancement is small. We show the effect in Fig.~\ref{fig:yfpeffect} (right) that, one needs smaller $Y_f$ to get relic density for non-zero $Y_{fi}^\prime$. The blue and red lines satisfy the relic density $\Omega_{DM}h^2=0.1198$. The blue line correspond to  $Y_{fi}^\prime=0$ whereas red line stand for the choice  $Y_{fi}^\prime=0.6$. The Higgs portal coupling $\kappa=0$, mixing angle $\cos\beta=0.995$ and masses $M_{E_1^\pm}=1500$ GeV and $M_{E_2^\pm}=3000$ are same in both cases.

It is also noteworthy that the non-zero Yukawa coupling $Y_{f2}$ (and $Y_{f1}=Y_{f3}=Y_f$) puts additional contribution to the processes $\Gamma(\mu \rightarrow e\gamma)$. and
throughout this analysis, we keep fixed $Y_{f2} = \mathcal{O}(10^{−3} )$. The parameters shown in Fig.~\ref{fig:relicall} are all allowed from the most stringent constraints of the flavour violating decay ~\cite{Baldini:2018nnn} BR$(\mu \rightarrow e\gamma) < 4.2 \times 10^{−13}$ at $90\%$ CL.
We also get negligibly small contributions to the electron as well as muon anomalous magnetic moment for the allowed parameters. These contributions can have positive and/or negative impacts depending upon the choice of parameters.
Possibly, with $M_S=1000$ GeV, $M_{E_1^\pm}=1500$ GeV and  $M_{E_2^\pm}=3000$ GeV, one needs $Y_{f2} =29.32$, to reproduce the present experimental data $\delta a_\mu=2.51\times10^{-11}$~\cite{Abi:2021gix}.
However, this choice of couplings and masses violate the LFV as well as perturbative limits $Y_{fi} > 4\pi$. Hence, the parameters allowed by relic density are also allowed by the recent electron as well as muon anomalous magnetic moment $g-2$ data at Fermilab~\cite{Abi:2021gix}.
However, the converse is not true in our model, as the region allowed
by electron and/or muon anomalous magnetic moment $g-2$ data violates the relic density,
LFV as well as perturbative bounds.

\subsection{FIMP dark matter}
We again want to remind the readers about the FIMPs. The main idea is that the dark matter sector gets populated through decay or annihilation of heavy particles until the number density of the corresponding heavy particle species become Boltzmann-suppressed. We need to solve Boltzmann equations that dictate the final relic abundance for the dark matter candidate $ S $.
One can easily calculate the decay width or annihilation of the heavy mother particles from the other particles. The coupling strengths are $\mathcal{O}(1)$, and we get ${\Gamma / H }>> 1$ for the mother particles.
Hence, all the mother particles (SM particles including heavy fermions) remain in the thermal equilibrium in the early Universe. Therefore we do not need to solve the Boltzmann equation for the evaluation of the mother particles, rather we have only solved the evaluation equation for the dark matter produces from the decay and annihilation of various mother particles.

This model can produce dark matter from the decay of the heavy fermions ($X_{Heavy}=E_1^\pm,E_2^\pm$ and $X_1^0$) and Higgs. 
It has already been noticed in the existing literature~\cite{Borah:2018gjk,Hall:2009bx,Biswas:2016bfo,Yaguna:2011qn,Bernal:2017kxu} that if the same couplings are involved in both decays as well as scattering processes, then the former has the dominant contribution to DM relic density over the latter one. The scattering processes have negligibly small contribution to DM relic density~\cite{Borah:2018gjk,Hall:2009bx,Biswas:2016bfo}. 
With reference to these past studies, we consider that the dark matter candidate is stable in our model and can produce only from the decay of the heavy vector fermions and Higgs.

The Boltzmann equation for the dark matter can be written as~\cite{Plehn:2017fdg,Hall:2009bx,Biswas:2016bfo},
{\small
	\begin{eqnarray}
		\hspace{-0.7cm} \frac{dn}{dt}+3 H n &=& -  \sum_i \mathcal{S}(X_{Heavy} \rightarrow   S    S , f_{\rm SM}  S  ),
		\label{eq:boltz1}
\end{eqnarray}}
where, $X_{Heavy}=E_1^\pm,E_2^\pm, X_1^0, H$ and $f_{\rm SM} $ is SM leptons.
Here, the decay-based source term $\mathcal{S}$ can be written as,
\bea
\mathcal{S} = \Gamma (X_{Heavy} \rightarrow   f_{\rm SM}   S , S    S ) \, \frac{K_1(\frac{m_{X_{Heavy}}}{T})}{K_2(\frac{m_{X_{Heavy}}}{T})} \, n^{eq}_{Heavy,i}
\eea
where, $K_{1,2}$ is themodified Bessel function of the first and second kind. For $x=\frac{m_{X_{Heavy}}}{T}$ and $Y=\frac{n}{T^3}$, the Boltzmann equation from eqn.~\eqref{eq:boltz1} now reads~\cite{Plehn:2017fdg},
\bea
\frac{dY(x)}{dx} = \sum_i \frac{g_{X_{Heavy}}}{2 \pi^2} \frac{\Gamma (X_{Heavy} \rightarrow  f_{\rm SM}   S , S    S ) }{H (x\approx 1)} x^3 K_1(x), 
\eea
where, $g_{X_{Heavy}}$ is the degrees of freedom of the heavy particle. We can integrate the dark matter production over the entire thermal history and look for the
final yield $Y(x_0)$ with the help of the appropriate integral~\cite{Plehn:2017fdg,Hall:2009bx,Biswas:2016bfo}as,
\bea
Y(x_0)=\frac{45 M_{Pl}}{6.64 \, \pi ^4 \, g^S \sqrt{g^\rho}} \,\sum_i  \frac{g_{X_{Heavy}}}{ M_{X_{Heavy,i}}^2} \Gamma (X_{Heavy,i} \rightarrow  f_{\rm SM}   S , S    S  ) \, \int_{x_{min}}^{x_{max}} x^3 K_{1} (x) dx,
\eea
 with, $g^S, \sqrt{g^\rho}$ are the effective numbers of degrees of freedom in the bath at the freeze-in temperature for the entropy, $s$, and energy density $\rho$.
Finally the relic density~\footnote{
We also crosscheck by using the output of {\tt FeynRules}~\cite{Alloul:2013bka} into {\tt micrOMEGAs}~\cite{Belanger:2018mqt} and get the same results.} ($M_S\equiv M_{DM}$) can be written as~\cite{Plehn:2017fdg,Hall:2009bx,Biswas:2016bfo},
\bea
\Omega h^2 &&= \frac{h^2}{3\, H_0^2\, M_{Pl}^2} \, \frac{ M_{ S }}{28} T_0^3 \, Y(x_0)\nn\\&&\approx 1.09 \times 10^{27} \, M_{ S }  \,\sum_i  \frac{g_{X_{Heavy,i}} \, \Gamma (X_{Heavy,i} \rightarrow  f_{\rm SM}   S , S    S  ) }{ M_{X_{Heavy,i}}^2} 
\label{eq:totalOmega}
\eea
\begin{figure}[h!]
	\begin{center}
		{\includegraphics[scale=.6]{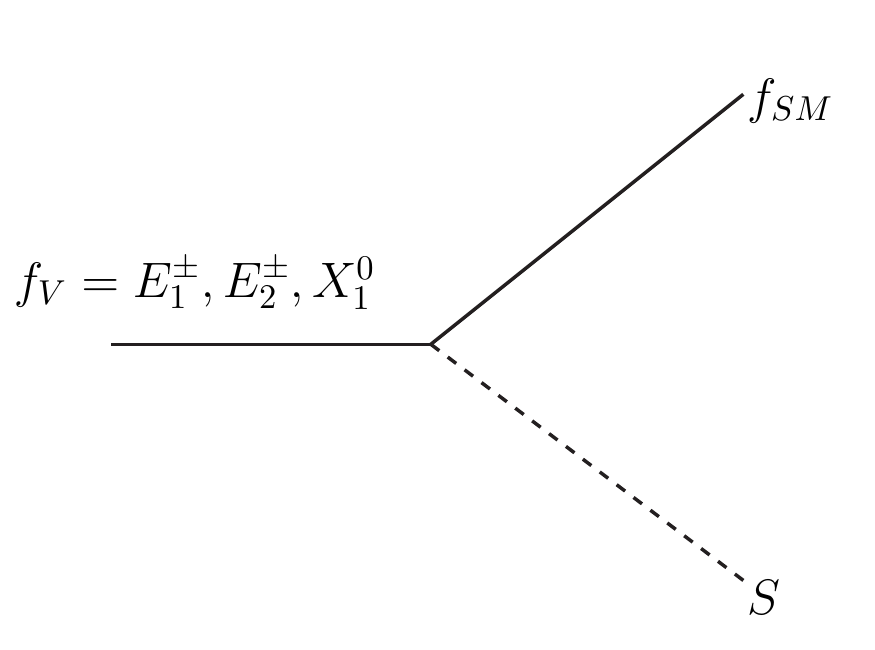}}
		{\includegraphics[scale=.6]{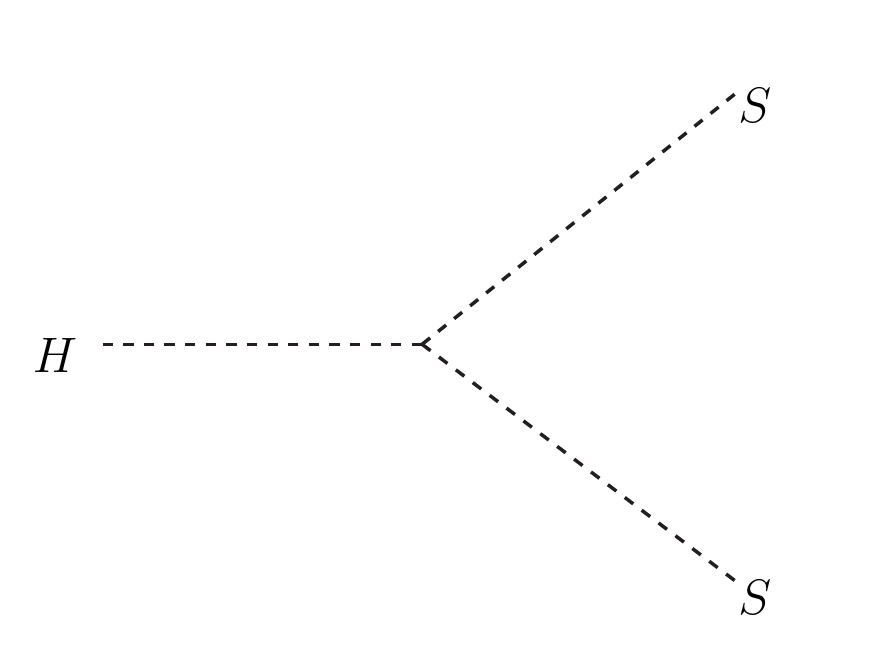}} 
		{\includegraphics[scale=.6]{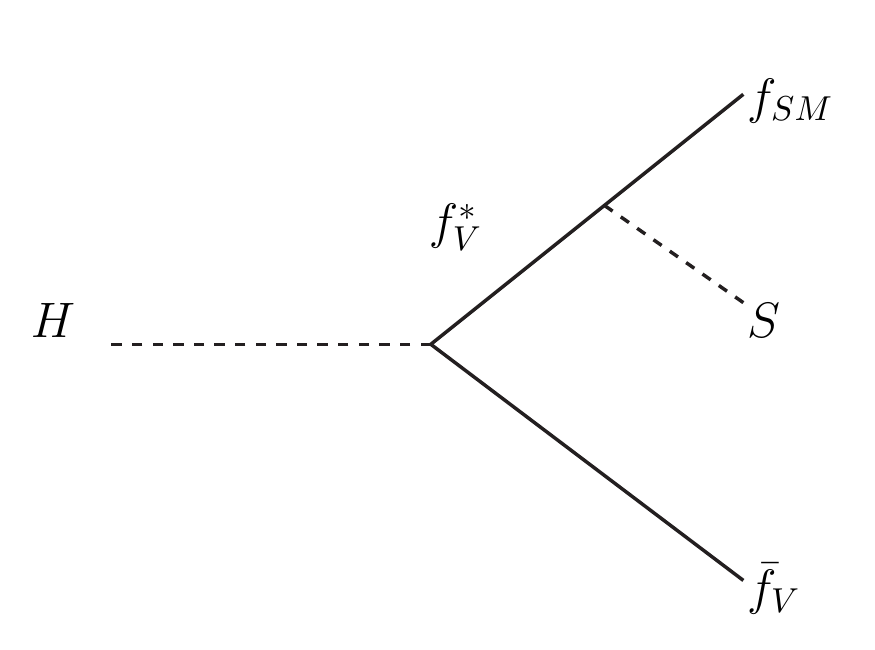}} 
		
		\caption{\it   \label{fig:decayHDD} \textit{ Dark matter production diagrams from the decay of the heavy particles contribute to the relic density.}}
	\end{center}
\end{figure}

We will now follow up the eqn.~\eqref{eq:totalOmega} to calculate the relic density. The main production diagrams from the decay of the heavy particles are shown in Fig.~\ref{fig:decayHDD}.
The last diagram is kinematically forbidden as we have considered $M_{f_V} \sim 1.5$ TeV. It is to be noted that, the heavy fermion mass of $M_{f_V} \sim 1.5$ TeV can lead to dilepton plus transverse missing energy signature $\gtrsim 3 \sigma$ at the  LHC for $\sqrt{s}=14$ TeV with an integrated luminosity $3000~{\rm fb^{-1}}$~\cite{Das:2020hpd}.
The additional four-body decay ($H \rightarrow \bar{f}_V^* f_V^*, f_V^* \rightarrow f_{SM} S \Rightarrow \bar{f}_{SM} f_{SM}\, SS$) diagram are suppressed by the heavy fermion propagator.
The partial decays of the heavy fermions and Higgs into the dark matter particle are  given by,
\bea
&&\Gamma(f_V \rightarrow  f_{\rm SM}   S ) = \frac{M_{f_V}}{8 \pi } \, |g_{f_V \, f_{SM} \,   S }|^2 \\
&&\Gamma(H \rightarrow   S    S ) = \frac{1}{32 \pi \,M_{H}} \, |g_{H  S    S }|^2 \, \left( 1 -  \frac{M_{ S }^2}{M_{H}^2}  \right)^{1 \over 2},
\label{eq:decayMF}
\eea
where, the coupling strengths are (see eqn.~\ref{eq:annivertex}) $
g_{HSS}= | \kappa v|$, $g_{SiE_1^\pm}= A_{fi}=  |(\cos\beta Y_{fi} + \sin\beta Y_{fi}^{\prime})|$, $g_{SiE_2^\pm}= B_{fi}= |(\sin\beta Y_{fi} + \cos\beta Y_{fi}^{\prime})|$ and $
g_{S\nu_i X_1^0}=C_{fi}= | Y_{fi}| , ~{\rm with}~i=e,\mu,\tau$.

Now we will discuss the numerical analysis for the FIMP dark matter region of the model parameter spaces. Let us first neglect the contribution from the decay  and annihilation from the scalar sector, i.e., $\kappa=0$, hence $\Gamma(H \rightarrow   S    S )=0$ and $\sigma({\rm SM~ particles}\rightarrow   S    S )=0$. It is to be noted that the mass of the dark matter can be set by the independent parameter $m_S$ (see eqn.~\eqref{eq:mass}).
We find that the $t$-channel annihilation processes $\sigma(f_V f_V \rightarrow  S   S)$ through $f_{\rm SM}$ propagator and $\sigma(f_{\rm SM} f_{\rm SM} \rightarrow  S   S)$ through $f_V$ propagator are also suppresses as compared to the decay $\Gamma(f_V \rightarrow  f_{\rm SM}   S )$ contributions.
\begin{figure}[h]
	\centering
	\includegraphics[scale=.30]{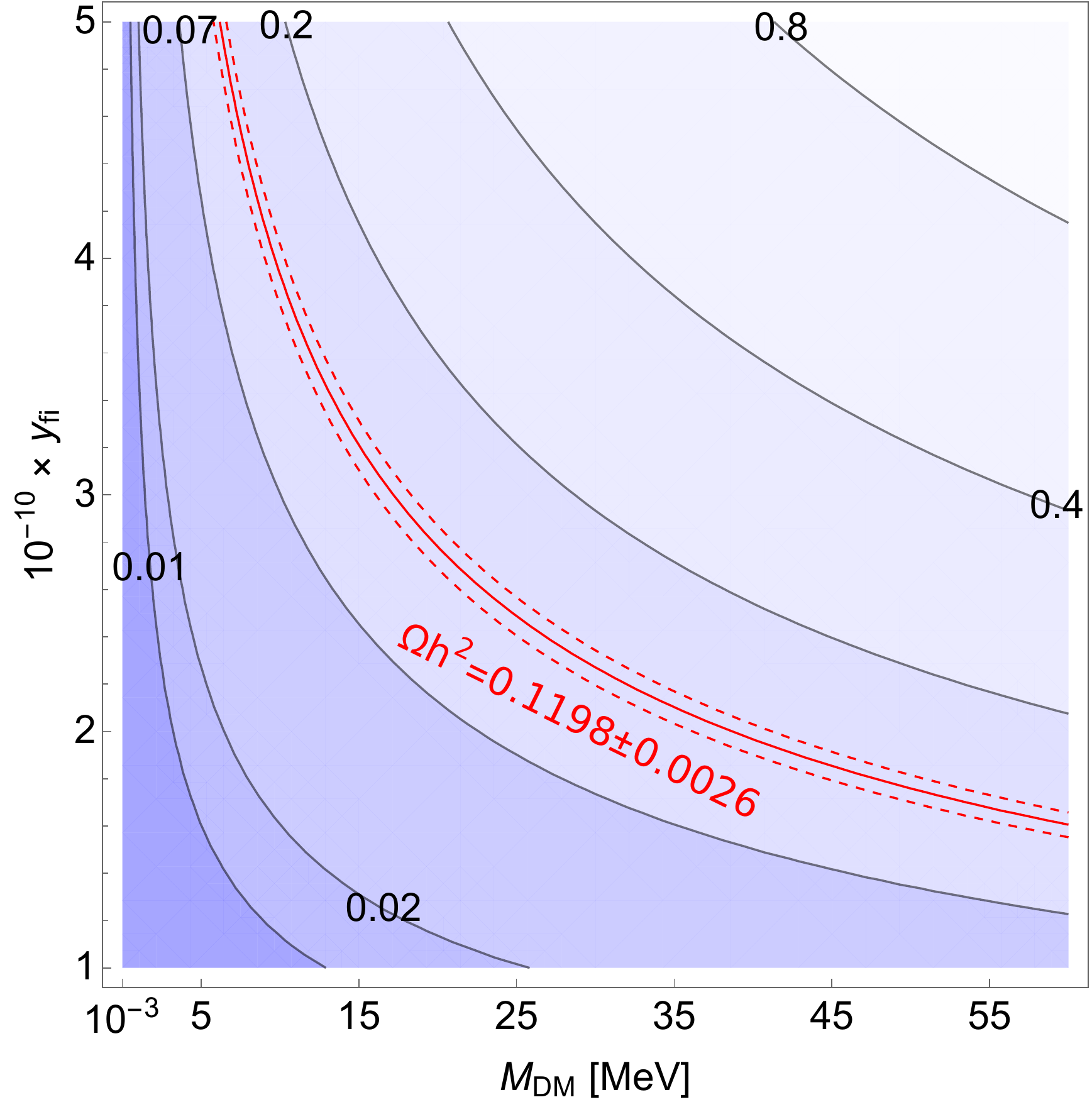}
	\includegraphics[scale=.30]{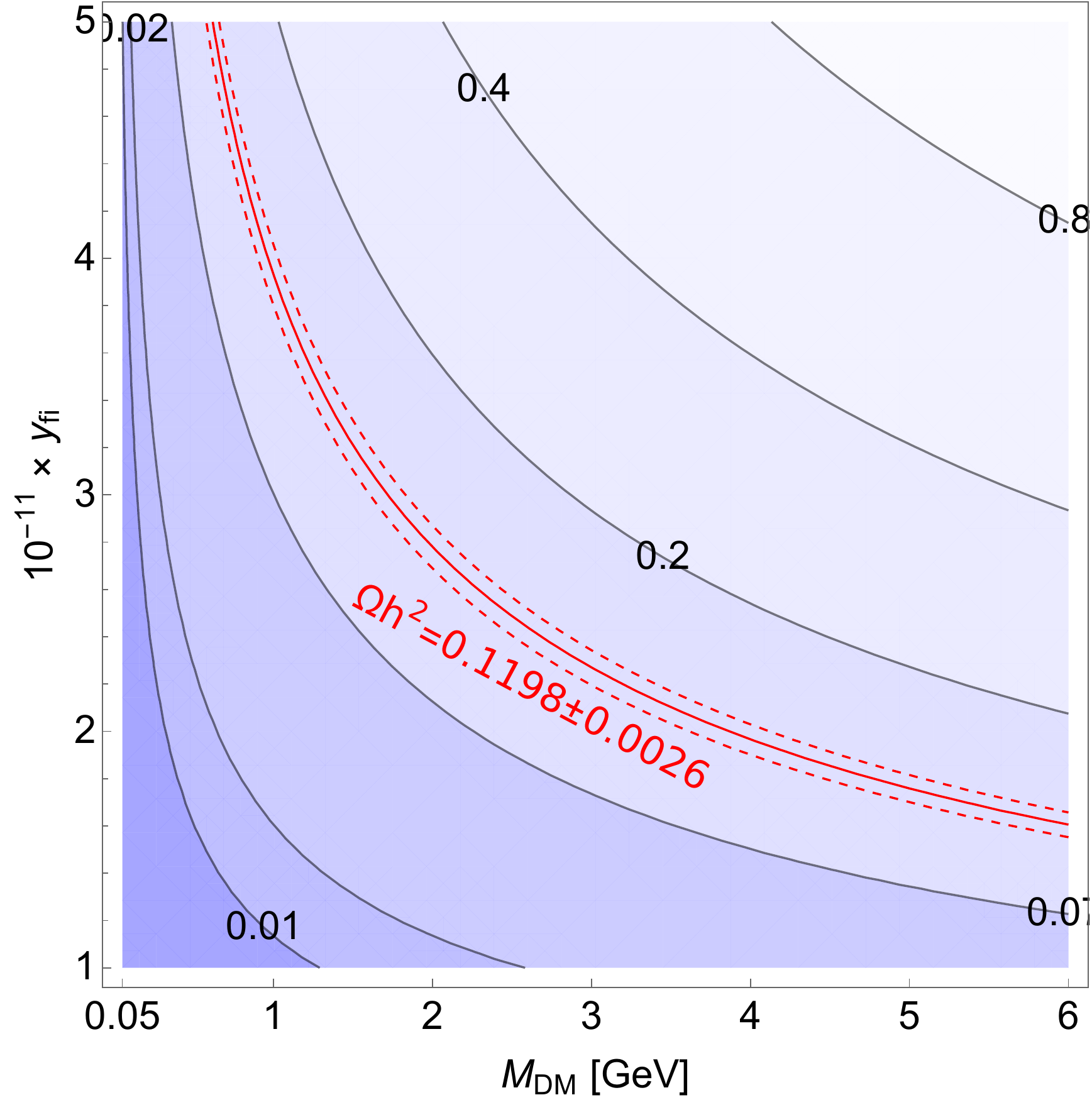}
	\includegraphics[scale=.310]{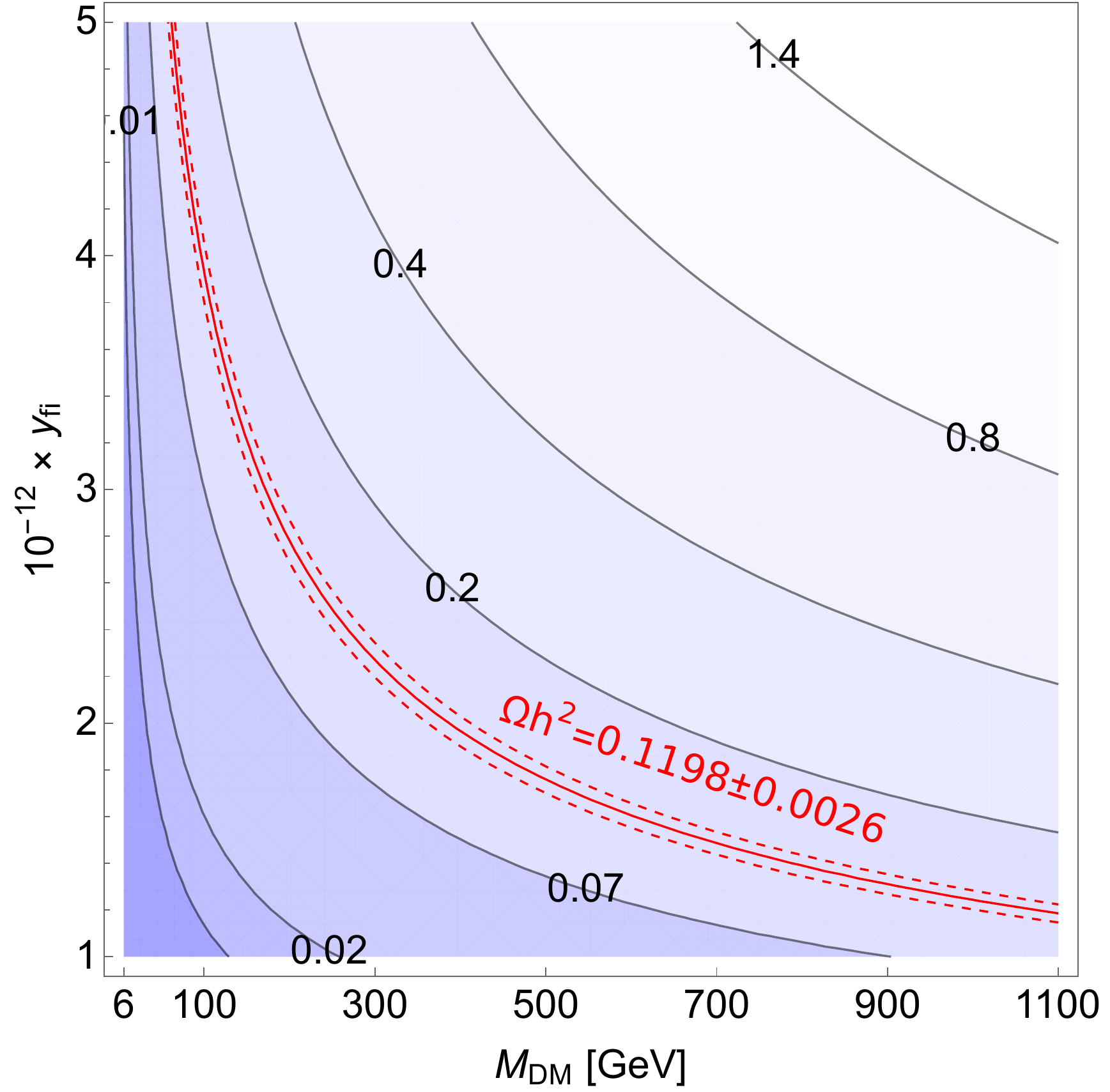}
	\caption{\it  The contour lines stand for the relic density in the new Yukawa coupling vs. dark matter mass plane. The red-lines indicate the relic density within the $3\sigma$ range. These region with $Y_{fi}<<1$ are also allowed by the recent LFV ~\cite{Baldini:2018nnn} $BR(\mu \rightarrow e\gamma) $ $< 4.2 \times 10^{−13}$ at $90\%$ CL, electron as well as muon anomalous magnetic moment $g-2$ data at Fermilab~\cite{Abi:2021gix}. The masses $M_{E_1^\pm}=1500$ GeV and $M_{E_2^\pm}=3000$ GeV and $\cos\beta=0.995$, hence $M_{X_1^0}=1514.9625$ GeV.}\label{Fig:relicFIMP1}
\end{figure} 
The variation of the new vector-like Yukawa coupling $Y_{fi}$ ($i=e,\mu,\tau$) with the mass of the dark matter are shown in Fig.~\ref{Fig:relicFIMP1}. 
In the meanwhile, we have neglected the effect from the interaction term $- \, Y_{fi}^{\prime }\, \overline{l}_{i,R} E_S S$ by considering $Y_{fi}^{\prime}=0$ in these plots. The dominant contributions ($>95 \%$) to the relic density mainly come from the decay of the fermions $E_1^\pm$ and $X_1^0$.
In both plots, the solid red line represents $\Omega h^2 = 0.1198$, and the red dashed lines correspond to the $3\sigma$ variation in $\Omega h^2$. The lighter region corresponds to higher values of $\Omega h^2$, which over close the universe and these regions strictly forbidden. The left plot is kept for the dark matter mass region $M_S\equiv M_{DM}=6$ MeV to $60$ MeV, whereas the right plot corresponds to the dark matter mass region  $60$ MeV to $6$ GeV. One can also get the exact relic density for the dark matter mass $\mathcal{O}(1)$ keV region, but it will create a problem for the structure formation.
 We calculate the free streaming length by following the formula given in equation (14) of the Ref.~\cite{Choi:2020kch},
\begin{equation}
\lambda_{FS}=\int_{a_{FS}}^1 \frac{da}{H_0 F(a)}\, \frac{<p_{DM}> a_{FS}}{\sqrt{ (<p_{DM}> a_{FS})^2 +  (m_{DM} a)^2 } },
\end{equation}		
where,	$F(a)=\sqrt{ \Omega_{rad,0} + a \Omega_{m,0}  + a^4 \Omega_{\Lambda,0} } $. We get $<p_{DM}> \approx \frac{M_H}{2}$ and 	$a_{FS}$ is calculated by equating the decay rate of the Higgs into two DM ($\Gamma(H \rightarrow SS)$) to the Hubble expansion rate $H\approx \frac{T^2}{M_{Planck}}$ during radiation dominated era and $H_0$ is the present Hubble expansion rate. The latest values of the cosmological parameters can be found in Ref.~\cite{Planck:2018vyg}.
We find the free streaming length larger than 10 Mpc; hence, we avoid showing these regions in this analysis. We also calculate the free streaming length~\cite{Choi:2020kch,Dev:2013yza} of the dark matter particles, and it is coming out to be less than $\mathcal{O}(100)$ kpc in the parameter space we have shown in both plots of Fig.~\ref{Fig:relicFIMP1}. Therefore, the dark matter still behave like a cold one~\cite{Dev:2013yza}.
For the heavier dark matter mass region, we need smaller Yukawa coupling as it compensate by the dark matter mass in the decay width. For example, we get the relic density in the right ballpark for $Y_{fi}= 1.244 \times 10^{-9}$ with $M_{DM} = 1 $ MeV and $Y_{fi}= 1.759 \times 10^{-10}$ with $M_{DM} = 50 $ MeV and we also get the relic density in the right ballpark for $Y_{fi}= 3.935 \times 10^{-11}$ with $M_{DM} = 1 $ GeV and $Y_{fi}= 1.759 \times 10^{-11}$ with $M_{DM} = 5 $ GeV. Hence, we get the relic density for $Y_{fi}\sim \mathcal{O}(10^{-10})$ through the Freeze-in scenario, and the contributions are tiny $\mathcal{O}(10^{-32})$ to the recent LFV $BR(\mu \rightarrow e\gamma) $ ~\cite{Baldini:2018nnn}, electron as well as muon anomalous magnetic moment~\cite{Abi:2021gix} $g-2$ data.

	We also change the mixing angle from $\cos\beta=0.995$ to  $\cos\beta=0.60$, keep fixed value of the masses $M_{E_1^\pm}=1500$ GeV and $M_{E_2^\pm}=3000$ GeV. We find a tiny change in the relic density; however, the plots in Fig.~\ref{Fig:relicFIMP1} remain almost the same. Remarkably, the relic density gets hugely modified for the WIMP dark matter scenario in the Freeze-out mechanism, and they are shown in Fig.~\ref{fig:yfpeffect}. The effect of the mixing angle is almost negligible in this Freeze-in mechanism where the dark matter gets produce from the decay of the heavy particles.
	For the fixed value of $\cos\beta=0.995$ and $M_{E_1^\pm}=1500$ GeV, one can also varied the mass 
	$M_{E_2^\pm}$, we find a very small effect in the relic density as the decay rate of $E_2^\pm$ is far smaller than the decay rate of $E_1^\pm$.

\begin{figure}[h]
	\centering
	\includegraphics[scale=.40]{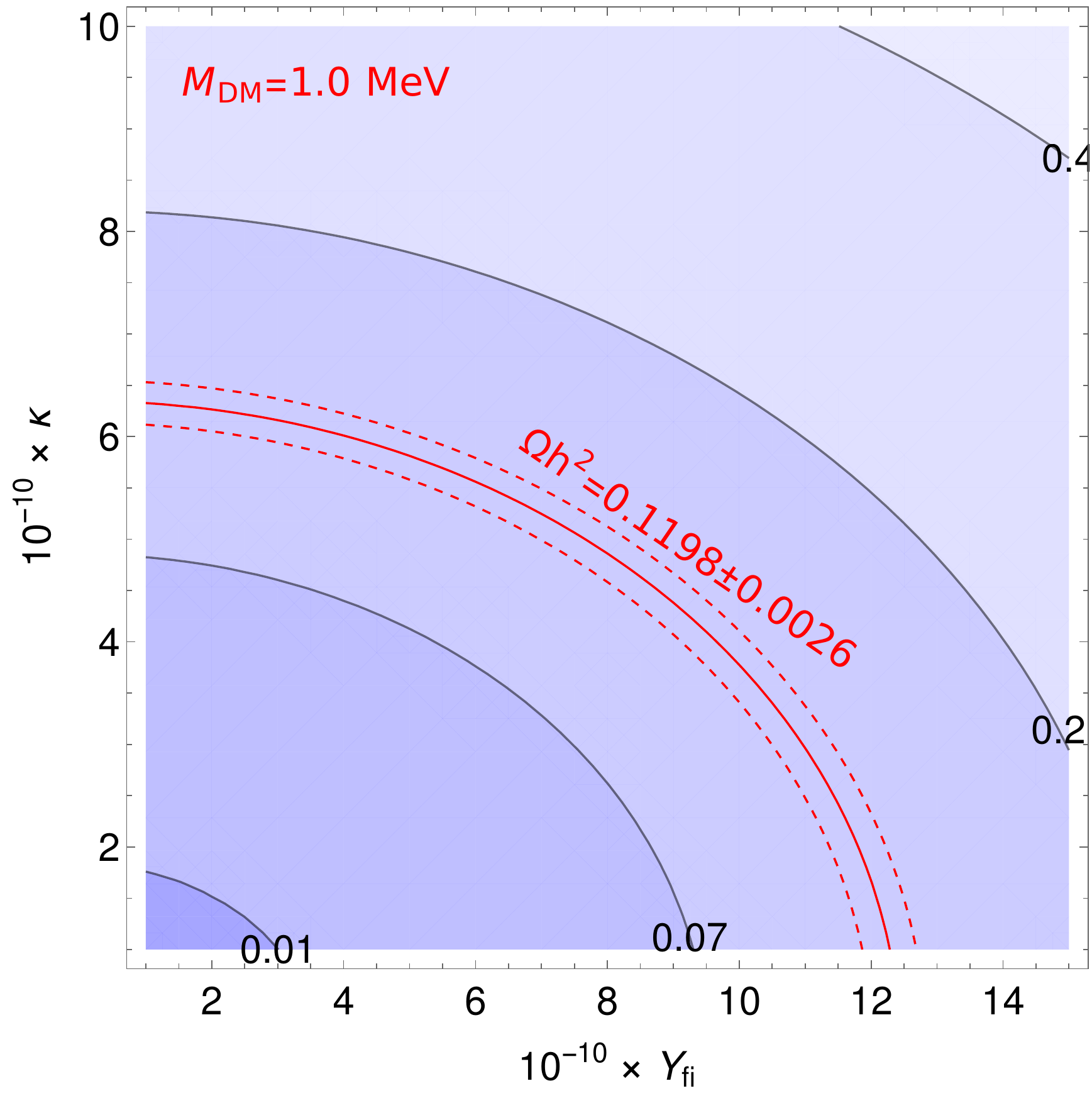}
	\includegraphics[scale=.40]{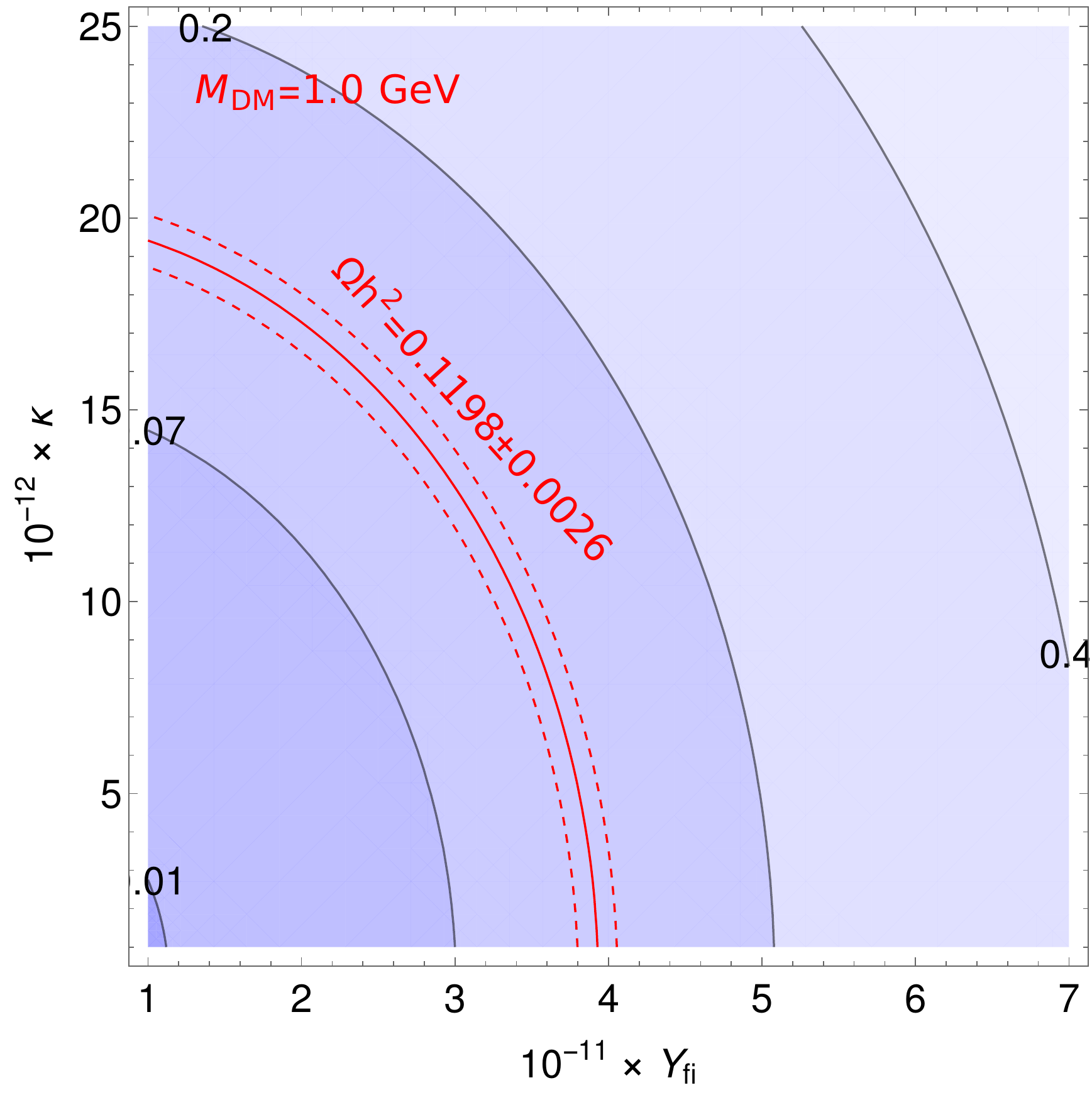}
	\caption{\it  The contour lines stand for the relic density in the new Yukawa coupling vs. dark matter mass plane. The red-lines indicate the relic density within the $3\sigma$ range. These region are also allowed by the recent LFV, electron as well as muon anomalous magnetic moment $g-2$ data. The masses $M_{E_1^\pm}=1500$ GeV and $M_{E_2^\pm}=3000$ GeV and $\cos\beta=0.995$, hence $M_{X_1^0}=1514.9625$ GeV. }\label{Fig:relicFIMP2}
\end{figure}

We now consider the contributions from the Higgs decay in the relic density. In this case, we still assume $Y_{fi}^{\prime}= 0$, to diminish the contributions from decay of $E_2^\pm$. The larger $\kappa$ increases the Higgs decays' contributions, whereas large values of $Y_{fi}$ increases the contributions from the vector-like fermions $E_1^\pm$ and $X_1^0$.
We show such variations for two different dark matter masses $M_{DM} = 1$ MeV and $1$ GeV respectively in Fig.~\ref{Fig:relicFIMP2}. The solid red line in both plots corresponds $\Omega h^2 = 0.1198$, and the red dashed lines represent the $3\sigma$ variation in $\Omega h^2$. The lighter region will over close the Universe. Therefore, very small $\kappa=\mathcal{O}(10^{-10})$ and $Y_{fi}=\mathcal{O}(10^{-10})$ are needed to get the exact relic density in the right ballpark. The additional fine-tune issues in these low dark matter mass regions are: (a) How could one get such a small dark matter mass? (b) How could one get such small couplings? One of the answers might be the Freeze-in mechanism for FIMP dark matter, and we need small coupling and mass to get the exact relic density. In the following collider section, we will discuss, whether it is possible to put additional constraints on DM mass from the indirect dark matter detection and/or from the collider experiments, if we have a tiny dark matter mass.

\begin{figure}[h]
	\centering
	\includegraphics[scale=.30]{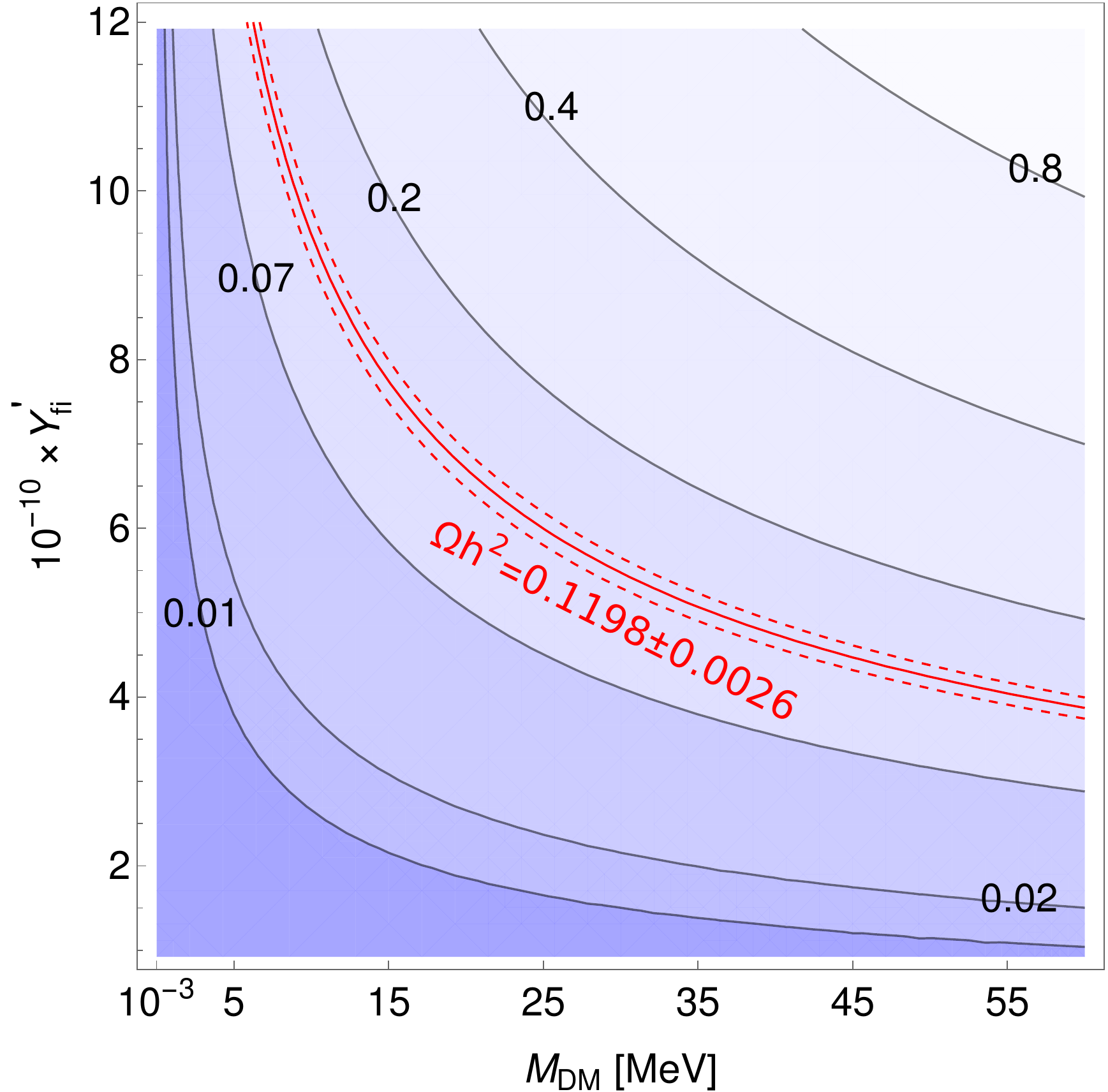}
	\includegraphics[scale=.30]{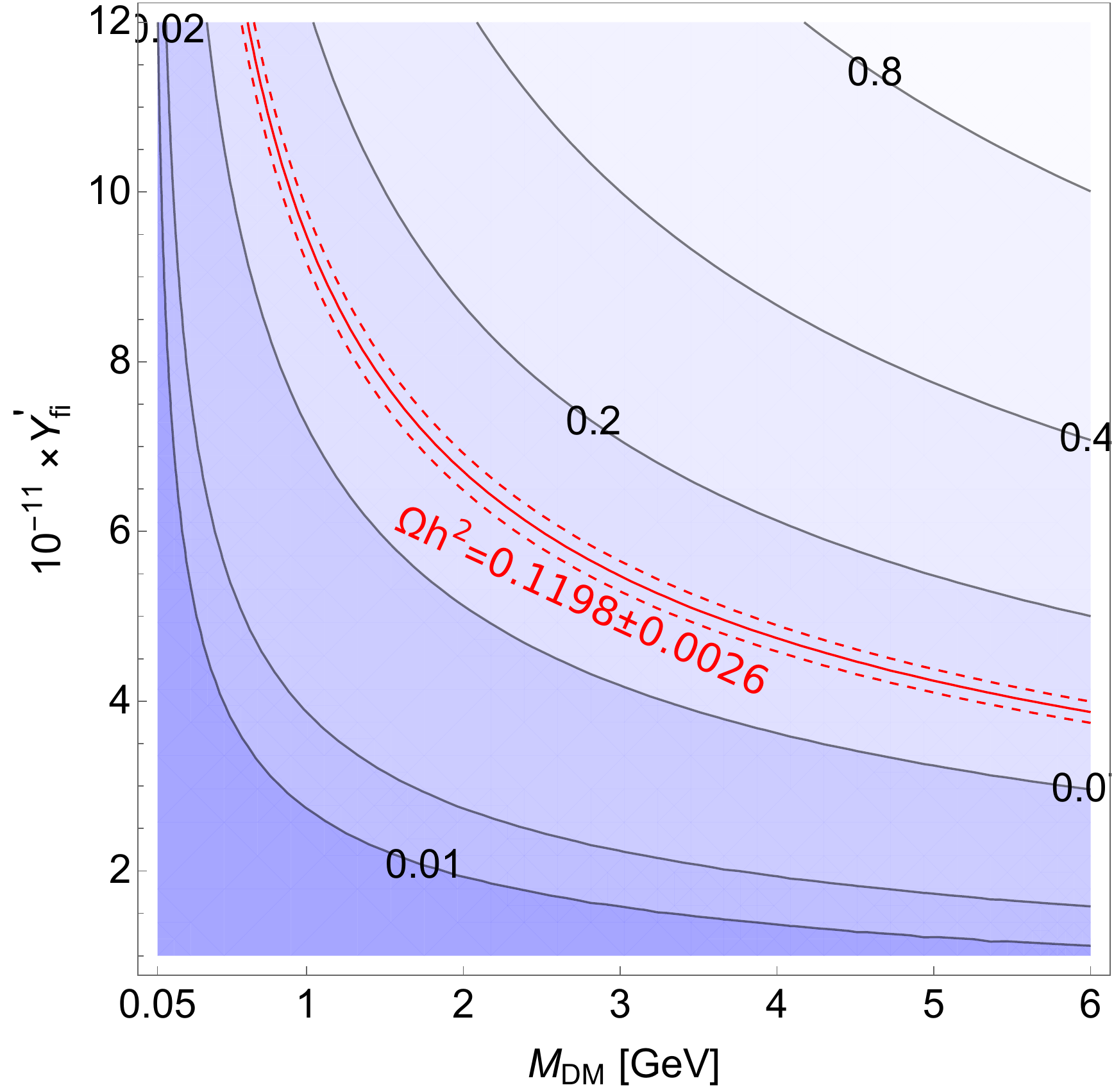}
	\includegraphics[scale=.31]{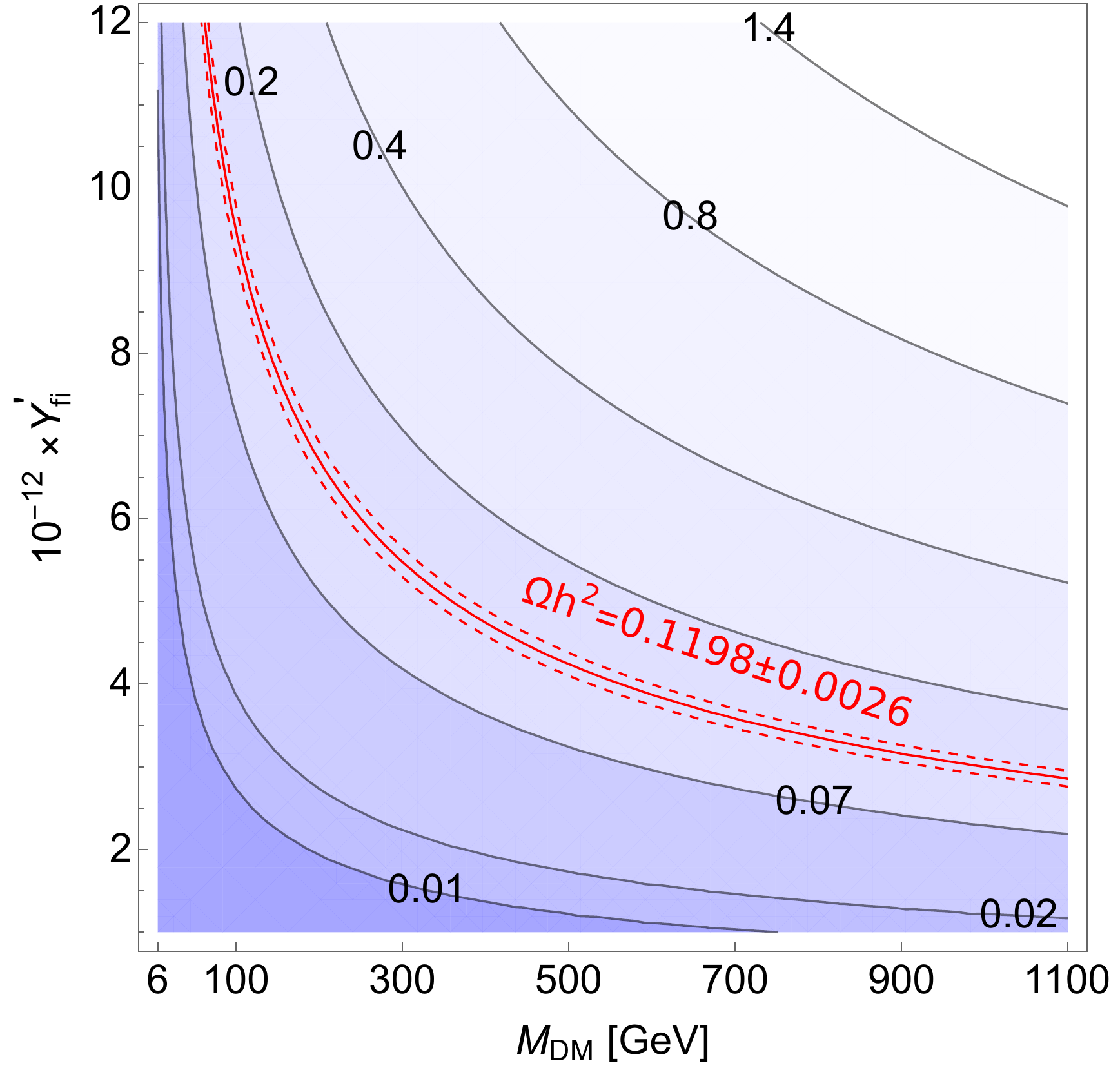}
	\caption{\it  The contour lines stand for the relic density in the new Yukawa coupling vs. dark matter mass plane. The red-lines indicate the relic density within the $3\sigma$ range. }\label{Fig:relicFIMP3}
\end{figure}

Now, we check the effect for the interaction term $Y_{fi}^{\prime}\bar{l}_{i,R}E_SS$ with $Y_{fi}^{\prime}\neq 0$. It will increase the contribution from the second charged fermion $E_2^\pm$. If we assume the $Y_{fi}=0$ and $\kappa=0$, then in the relic density, the dominant contribution come from the decay of this charged fermion $E_2^\pm$. In  Fig.~\ref{Fig:relicFIMP3}, we show the contour regions for the relic density in $Y_{fi}^\prime$ $vs.$ dark matter mass (both MeV and GeV region) plane. The red lines within the $3\sigma$ range. The contribution from the other charged fermion $E_1^\pm$ is less than one percent for this choice of $\cos\beta=0.995$.
%
%
%
%
%

\begin{figure}[h]
	\centering
	\includegraphics[scale=.41]{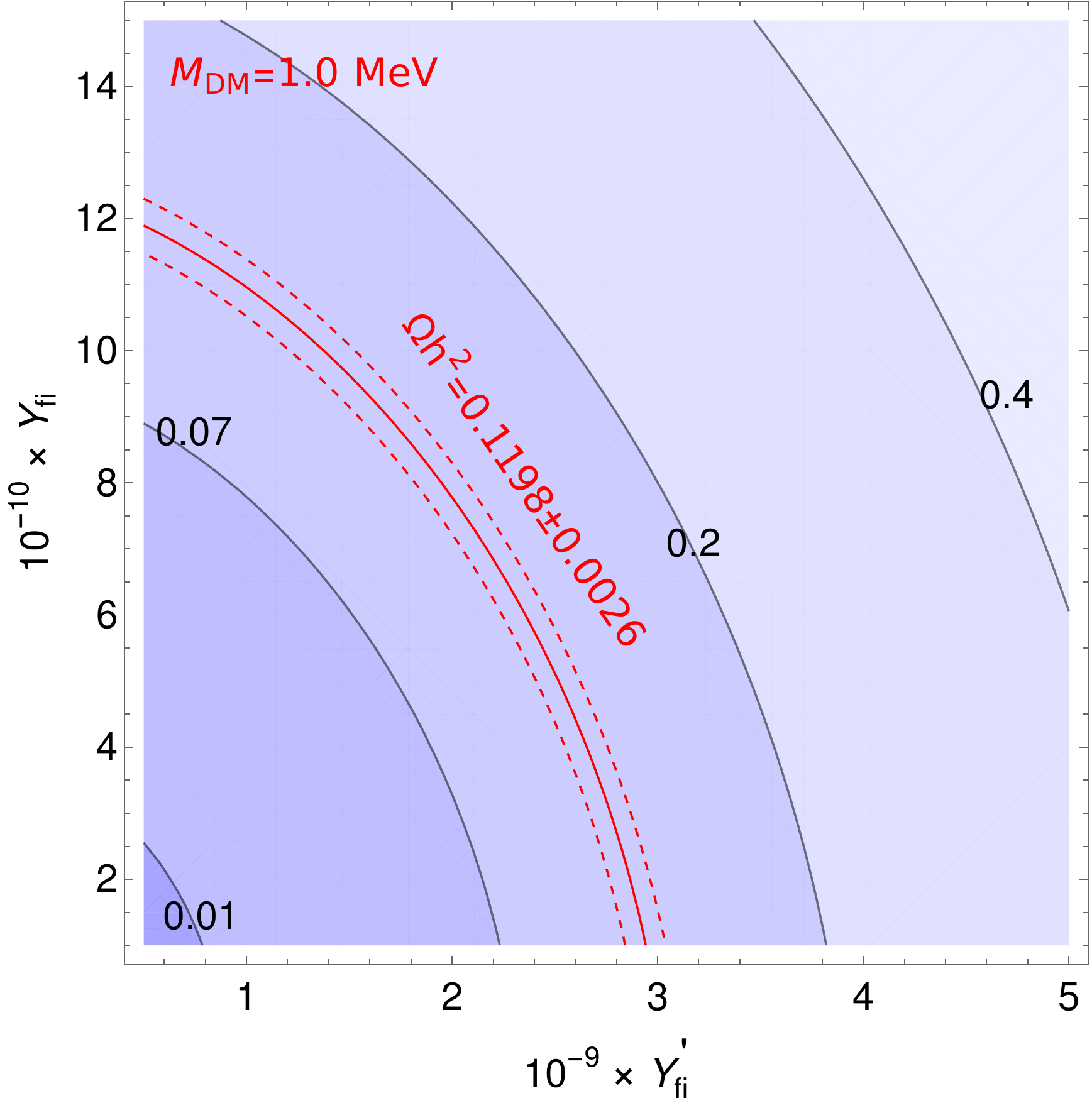}
	\includegraphics[scale=.40]{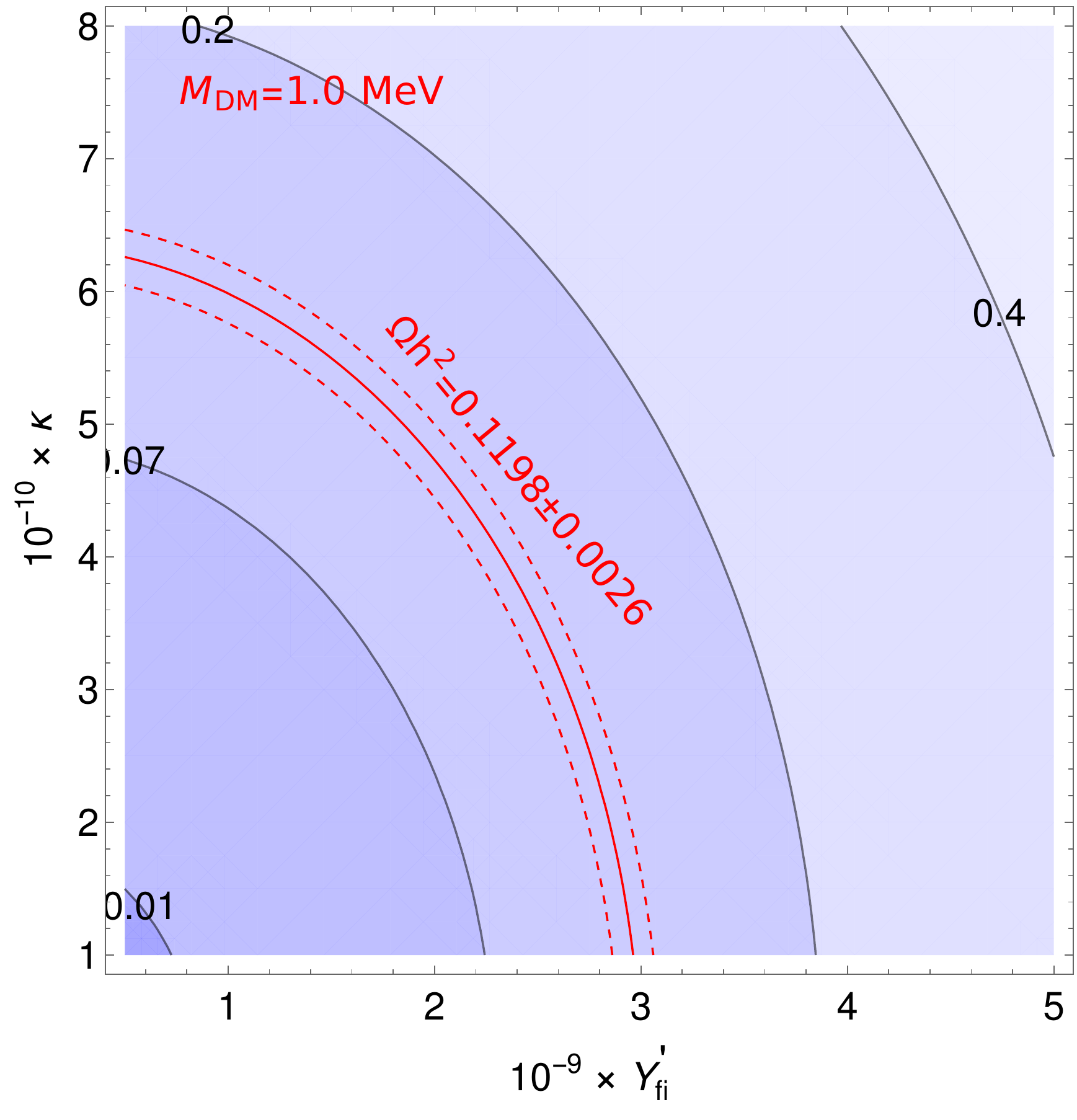}
	\caption{\it  The contour lines stand for the relic density in the new Yukawa coupling ($Y_{fi}^{\prime}$) vs. the old Yukawa coupling ($Y_{fi}$) and $\kappa$ respectively. The red-lines indicate the relic density within the $3\sigma$ range. }\label{Fig:relicFIMP4}
\end{figure}

%
%
In the presence of $Y_{fi}$ and $\kappa$, one can have additional contributions to the relic density from the decay of charged fermion $E_1^\pm$ and Higgs scalar. In these two contour plot in  Fig.~\ref{Fig:relicFIMP4}, we keep x-axis reserve for $Y_{fi}^{\prime}$ and y-axis for $Y_{fi}$ and $\kappa$ respectively. We neglect the contribution from the Higgs decay in the left plot of Fig.~\ref{Fig:relicFIMP4}, i.e., $\kappa=0$ whereas the $Y_{fl}=0$ in right plot. We consider fixed dark matter mass at $1$ MeV and we keep $\cos\beta=0.995$, $M_{E_1^\pm}=1500$ GeV and $M_{E_2^\pm}=3000$ GeV.  
%
%

\section{Indirect Detection limits}
The normal gamma-ray limits from the Fermi-LAT do not apply for  the mass $<\mathcal{O}(100)$ MeV, although the data from several other satellites such as HEAO-1~\cite{Gruber:1999yr}, INTEGRAL~\cite{Bouchet:2008rp}, COMPTEL~\cite{kappadath}, EGRET~\cite{Strong:2004de} are extremely sensitive to photons with energies well below a $\mathcal{O}(100)$ MeV. 
For a DM  annihilation spectrum $dN_\gamma/dE$, and a galactic DM density profile $\rho (r)$, the galactic contribution to the differential photon flux per unit energy is given by~\cite{Essig:2013goa},
\bea
{d\Phi_{\gamma,G} \over dE d\Omega } = \frac{ r_\odot \, \rho_\odot \, \Gamma_{A}}{ 4\pi \,  2^{\alpha-1}\, M_{ DM } } \, J_{D,A}\, {dN_\gamma \over dE},
\eea
where, $\Gamma_{A}=\frac{\rho_\odot}{M_{DM}}<\sigma v_{DM}>$, $r_\odot \sim 8.5$ kpc is the Sun's distance from the Galactic center and $\rho_\odot=0.3\,{\rm GeV\,cm^{-3}}$ is the local DM density. $\alpha = 1 (2)$ for DM decays (annihilations) and 
$
J_{A} = \int_{l.o.s} \frac{ds}{r_\odot}\, \left( \frac{\rho(s)}{\rho_\odot} \right)^\alpha
$
is a dimensionless quantity that describes the density of decays or annihilations along the line-of-sight (l.o.s.), and $\rho(s)=\rho(r,l,b)$ can be found in~\cite{Navarro:1996gj}. 
The extra-galactic photon spectrum arising from dark matter decays; thus, the contribution is zero here.
The $E^2\,{d\Phi_{\gamma,Tot} \over dE d\Omega } $ vs. $E$ limits are shown in Fig. 1 of the Ref.~\cite{Essig:2013goa}. We find the average dark matter annihilation cross-section $<\sigma v_{DM}> \approx 10^{-64}~{\rm cm^3/s}$ in the exact relic density region. We have found that to explain above experimental data, we need a very large $\kappa\sim Y_{fl}\sim 25$ which violates perturbativity.
Hence, the regions considered in this analysis are allowed from these experimental limits.
\section{Direct detection at Collider as charged track}\label{sec:collidr}
As we have mentioned that we need a tiny Higgs portal coupling $\kappa\sim 10^{-10}$ to get the exact relic density; hence, the direct detection limit (e.g., XENON-IT limit~\cite{Aprile:2018dbl}) is not applicable for the FIMP dark matter. 
The detailed collider bounds have been discussed in our previous paper~\cite{Das:2020hpd} for the same model in the context of WIMP scenario.
The authors of the Ref.~\cite{No:2019gvl} have discussed the method to prob the FIMP dark matter in the recent LHC collider with high integrated luminosity using the MATHUSLA surface detector. A charged track can be obtained in this model due to the vector fermions' decay into SM fermion and dark matter candidate at the collider. In this model, the length travelled by the charged fermion $E_{1,2}^\pm$ before its decay. 
\beq
c \tau_{E_{1,2}^\pm} = \frac{ 4.9627 \times 10^{-13}}{g_{fV}^2} \left( \frac{1 ~\rm [GeV]}{M_{E_{1,2}^\pm}} \right) ~{\rm cm}.
\eeq

Can we get enough events from this charged track? It mainly depends on the production cross-section $\sigma^{\rm LHC}_{\sqrt{s}}$ of the mother particle (vector fermion or Higgs) and luminosity $\mathcal{L}$ at the detector. The number of events at the LHC is calculated in Ref.~\cite{No:2019gvl}, and it is given by
\begin{equation}
	N_{events} = \sigma^{\rm LHC}_{\sqrt{s}} \, \mathcal{L}  \, \int P^{\rm MATH}_{\rm Decay},~~~~~{\rm with}~~~P^{\rm MATH}_{\rm Decay}=0.05(e^{-\frac{L_a}{\beta \, c \, \tau_{f_{MF}}}} - e^{-\frac{L_b}{\beta \, c \, \tau_{f_{MF}}}}),
\end{equation}
The author of Ref.~\cite{No:2019gvl}, find the number of events $N_{events}\geq 3$ for $\sqrt{s}=13$ TeV with an integrated luminosity $\mathcal{L}=3000 ~{\rm fb^{-1}}$ using their model parameter spaces. They showed that the MATHUSLA100/200 detector could detect these mother particle up to   $1$ TeV. The dominant production of the vector fermions come through the Drell-Yan processes.

In this model, we find the production cross-section of the vector like charged fermions with mass $M_{E_1^\pm}=1500$ GeV is $6.53\times 10^{-3}$ fb~\cite{Das:2020hpd}.
Hence we need large luminosity and/or energy to get a significant event at the present MATHUSLA100/200 surface detector. We find $N_{events}>3$ at 14 TeV LHC with an integrated luminosity $\mathcal{L}=10^6 ~{\rm fb^{-1}}$. The 14 TeV HL-LHC will collect data around $3000$ $fb^{-1}$, so this search will not be effective. We need wait for the $100$ TeV with high luminosity collider.

\section{Conclusion}\label{conc}
In this work, we study the possibility of dark matter in an extended singlet scalar model.
The structure of the model shown here uses a minimum number of particle content. This model contains two additional vector-like charged fermion and a neutral fermion along with a real singlet scalar field. In the previous study~\cite{Das:2020hpd}, we have added one extra vector-like fermion doublet in the model to complete the neutrino framework. We also see that the additional heavy vector-like fermions from this doublet do not alter the dark matter phenomenology as the mixing was taken to be very small. However, the neutrino sector is skipped in this work and rather focus on the WIMP and FIMP dark matter analysis. 

We revisited the dark matter analysis through a Freeze-out mechanism considering the collider bound as discussed in Ref.~\cite{Das:2020hpd}. Here, we show a broad region of the WIMP dark matter parameter spaces which satisfy the relic density at the right ballpark. We choose different dark matter parameter spaces to get the new allowed region from the relic density through the Freeze-out mechanism.
In the presence of vector-like fermions, one can get the correct relic density via co-annihilation, or one may have the interaction term such that the dark matter can annihilate into SM particles through additional cross-channel, $t$- and $u$-channels. The destructive and/or constructive interference among these channels helps to modify the effective average dark matter annihilation cross-section and provide the exact relic density in this model.

This study also focuses on the low dark matter parameter spaces where the dark matter density can explain through the Freeze-in mechanism. As we know, one cannot get the same viable dark matter parameter space in the low dark matter region as it will either violate the relic density, direct detection constraints or the perturbative limit. Interestingly, in this study, we showed that a single model could explain the low and high dark matter mass region allowed from all the other phenomenological constraints. 
If, in the future, we able to get any signature at a very low or high dark matter mass region (keV-TeV mass region), our present (including the previous) study could help in estimating a better parameter space.

A few comments on the recent muon anomalous magnetic moment experimental results are made in this model. 
We see that both the charged particles and the dark matter candidate played a crucial role in providing additional contributions.
It was found that one can explain the experimental data throughout the parameter spaces. However, the lepton violating decay channels, perturbative and unitary limits put stringent constraints on the parameter spaces.
The dark matter mass region allowed by relic density through Freeze-in and Freeze-out mechanisms is also allowed by the recent muon $g-2$ data at Fermilab. However, the converse is not valid in our model, as the region allowed by muon $g-2$ data violates the relic density bound.
We also perform the collider analysis to search the new particles in the FIMP like scenario in the context of the same 14 TeV LHC experiments with the MATHUSLA100/200 detector. 
A charged track can be obtained due to the decay of the heavy charged fermion $E_{1,2}^\pm$ into SM fermion and dark matter. One can get events larger than $N=3$, when the LHC operated with integrated luminosity $\mathcal{L}=10^6 ~{\rm fb^{-1}}$.

\section{Acknowledgements}
The research work of P.D. and  M.K.D. is supported by the Department of Science and Technology, Government of India under the project grant EMR/2017/001436.
NK would like to thank to Prof. Dilip Kumar Ghosh for his support at IACS. This project has also received funding/support from the European Union’s Horizon 2020 research and innovation programme under the Marie Skłodowska -Curie grant agreement No 860881-HIDDeN”.

\providecommand{\href}[2]{#2}\begingroup\raggedright\endgroup
\end{document}